%% file: main-astro-ph.tex
\documentclass[preprint2]{aastex}

\usepackage[figuresright]{rotating}
\usepackage{hyperref}
\usepackage{breakurl}

\slugcomment{Accepted in JQSRT}

\shorttitle{Contributions of artificial lighting sources on light pollution in Hong Kong}
\shortauthors{J. C. S. Pun et al.}

\begin{document}


\title{Contributions of artificial lighting sources on light pollution in Hong Kong measured through a night sky brightness monitoring
network}


\author{Chun Shing Jason Pun\altaffilmark{1}}
\email{jcspun@hku.hk}


\author{Chu Wing So\altaffilmark{1}}
\email{socw@hku.hk}


\author{Wai Yan Leung\altaffilmark{1}}
\email{yanyan12@hku.hk}

\and

\author{Chung Fai Wong}


\altaffiltext{1}{Department of Physics, The University of Hong Kong, Pokfulam Road, Hong Kong}


\begin{abstract}
\input{abstract}
\end{abstract}


\keywords{light pollution, night sky brightness, skyglow, Moon radiation, urban lighting}


\section{Introduction} \label{sec:introduction}
\input{introduction}

\section{Methodology} \label{sec:methodology}
\input{methodology}

\section{Night Sky Brightness in Hong Kong} \label{sec:results_analysis}
\input{results-analysis}
\section{Discussion} \label{sec:discussions}
\input{discussions}

\section*{Acknowledgments}
\input{acknowledge}



\bibliographystyle{aa}
\bibliography{AAstyleforJQSRT}



\end{document}

%% file: abstract.tex


Light pollution is a form of environmental degradation in which excessive artificial outdoor lighting, such as street lamps, neon signs, and illuminated signboards, affects the natural environment and the ecosystem. Poorly designed outdoor lighting not only wastes energy, money, and valuable Earth resources, but also robs us of our beautiful night sky. Effects of light pollution on the night sky can be evaluated by the skyglow caused by these artificial lighting sources, through measurements of the night sky brightness (NSB). The \textit{Hong Kong Night Sky Brightness Monitoring Network} (NSN) was established to monitor in detail the conditions of light pollution in Hong Kong. Monitoring stations were set up throughout the city covering a wide range of urban and rural settings to continuously measure the variations of the NSB. Over 4.6 million night sky measurements were collected from 18 distinct locations between May 2010 and March 2013. This huge dataset, over two thousand times larger than our previous survey \citep{pun:2012}, forms the backbone for studies of the temporal and geographical variations of this environmental parameter and its correlation with various natural and artificial factors. The concepts and methodology of the NSN were presented here, together with an analysis of the overall night sky conditions in Hong Kong. The average NSB in Hong Kong, excluding data affected by the Moon, was 16.8 mag~arcsec$^{-2}$, or 82 times brighter than the dark site standard established by the International Astronomical Union (IAU) \citep{smith:1979}. The urban night sky was on average 15 times brighter than that in a rural location, firmly establishing the effects of artificial lighting sources on the night sky. \\
\\

%% file: introduction.tex
Outdoor lighting is an indispensable element of modern civilized societies for safety, recreation, and decorating purposes. However, poorly designed outdoor lighting systems and excessive illumination levels can lead to light pollution \citep{smith:2009,starlight:2007}. The scattering of artificial light by cloud, aerosol, and pollutants such as suspended particulates in the atmosphere spread the effects to distances beyond the position of the lighting source and can brighten the entire night sky \citep{benn:newastro}. Light pollution is a form of environmental degradation in which excessive artificial outdoor lighting affect the natural environment and the ecosystem. It not only represents a waste of energy, money, and valuable Earth resources, but also indirectly contributes to the global environmental problems. Last but not least, the skyglow due to these artificial lighting sources leads to the degradation of the quality of night sky and reduces the number of observable stars, robbing us the beautiful night sky on dark nights when it would otherwise be visible.  


The energy wasted by artificial lighting sources can be directly monitored by night-time images of the Earth. Satellite imagery taken from the US Air Force Defence Meteorological Satellite Program (DMSP) Operational Linescan System (OLS)\footnote{\url{http://ngdc.noaa.gov/eog/}} sampled large areas of landmass at a moderate spatial resolution ($\sim$~1~km), making studies of the extent and degree of night time radiation over a metropolis or even globally possible \citep{kyba:2013,small:2013,cinzano:2004,cinzano:2001altlas}. 
The night-time photographs taken by astronauts onboard the International Space Station (ISS)\footnote{\url{http://eol.jsc.nasa.gov/}} provided even higher spatial resolution (at $\sim$~6~m per pixel) for selected locations on Earth. While the light intensity on the ISS photograph of Hong Kong was found to be positively correlated with population density in general \citep{liu:2011}, the scarcity of these photographs was insufficient for studies of nightly variations of light pollution in the city. 

The extent of light pollution can also be monitored by studying the night sky brightness (referred to as NSB hereafter). The NSB is a combination of the natural sky glow due to celestial objects (Sun, Moon, planets, stars, Milky Way, galaxies, etc.), and the sky glow due to direct or reflected light from artificial lighting sources. Apart from the effects of the Sun and the Moon, the chief contributor to the NSB for a highly populated metropolitan city is the artificial lighting source. The level of the NSB had mostly been studied at professional astronomical observatories using the traditional astronomical technique of photometry~--- the amount of light detected in star-free regions on the CCD images was extracted to estimate the NSB at different wavelength bands \citep{marco:2009,IAO:2008,patat:2008,sanchez:2007,krisciunas:2007,taylor:2004,liu:2003}. 
This kind of observation helped astronomers to reveal potential light pollution threats to the observatory and aid in their search for a new potential dark observing location. These observations would normally require skilled personnels using delicate experimental setup (telescope/lens/camera/mounting). The geographical and temporal coverages of those studies were usually limited (limited to a single site, and only a relatively few observations per night). 
 


Large-scale surveys of the sky conditions had also been carried out by campaigns which recruited ordinary people to conduct visual studies of the night sky following simple procedures with minimal technical expertise required, such as the 
\textit{GLOBE at Night}\footnote{\url{http://www.globeatnight.org/}} and the 
\textit{Great World Wide Star Count}\footnote{\url{http://www.windows2universe.org/citizen\_science/starcount/}} projects. Without the need for specialized or expensive equipment, this kind of citizen-science project encouraged a large number of people from around the world to take part in studies of light pollution, yielding results with broad geographical coverage and high spatial resolution if the number of observers were large enough. Furthermore, many observations were conducted near places where the participants lived, such as city centers or suburban regions, or at places that they could gain access to, such as country parks. Recent comparison of a subset of the \textit{GLOBE at Night} results with the night-time satellite images illustrated that this kind of study has huge potential for global scale studies of the night sky \citep{kyba:2013}. 

\begin{figure*}[ht]
\begin{center}
\includegraphics[width=16cm]{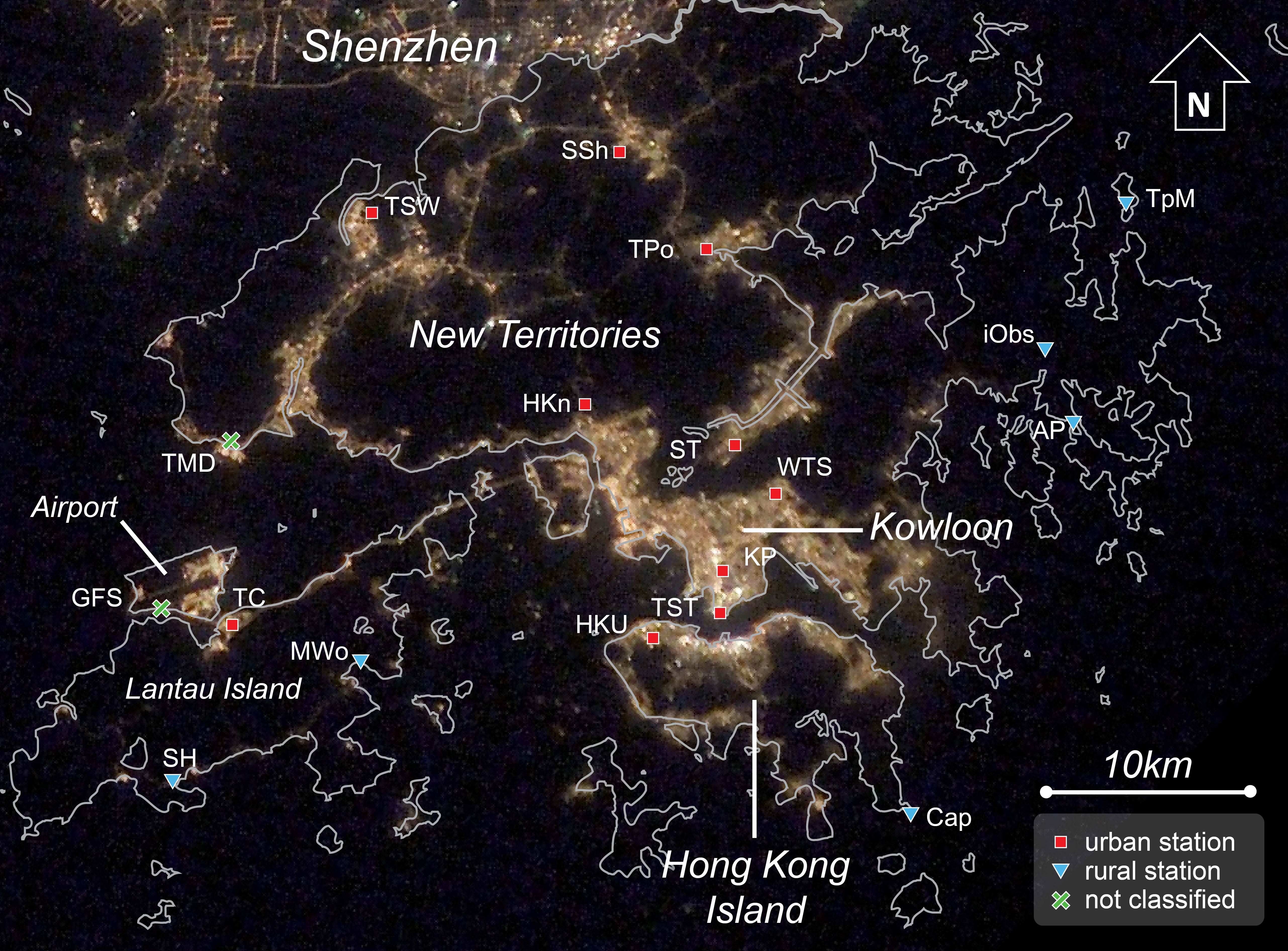}
\caption{Image taken from the International Space Station (ISS) when it flew over Hong Kong in orbit at altitude of $\sim$~300~km. The image was taken at 00:51 (UTC+8) on 11 March 2003 local time. Landmass boundaries and locations of NSN monitoring stations obtained from \textit{Google Map} are overlaid on the image. Refer to Table~\ref{tab:station-list} for station codes and Section~\ref{subsec:obssite} for details. (credit: Image Science and Analysis Laboratory, NASA-Johnson Space Center, \textit{The Gateway to Astronaut Photography of Earth})
\label{fig:station-map}}
\end{center}
\end{figure*}

Hong Kong is a populous metropolitan city (mid-2012 population 7,154,600 \citep{mid2012pop}) with a high population density (6,620 per km$^2$), known for its spectacular night lights. The mountainous city has a complex geographical landscape, leading to a short supply of habitable land. 
A photograph taken from the ISS in March 2003 (Figure~\ref{fig:station-map}) revealed the huge amount of upward shooting light at night. As seen in the figure, there were substantial variations in the level of night-time emission from the city, mostly due to differences in population and land utilization. It seems obvious that the degree of light pollution in Hong Kong should be strongly dependent on human activity, in particular, how and where people use external lighting. 

The availability of low-cost light sensors originally targeted for the astronomical community allowed for detailed and comprehensive studies of the NSB.
Between 2007 and 2009, we conducted a citizen-science survey of light pollution in Hong Kong by inviting students, astronomy enthusiasts, and campsite employers to measure the NSB using one such device, the Sky Quality Meter (SQM)~\citep{pun:2012}. Compared to projects such as \textit{GLOBE at Night}, the use of a standard measuring device in this study reduced the uncertainties in night sky measurements due to variations in observers' eyesight and experience. From the over 2,000 measurements taken at almost 200 locations by over 170 volunteers, it was concluded that light pollution in Hong Kong is severe, with large brightness contrast between the observed urban versus rural locations. Moreover, later night skies (at 23:30, local time (UTC+8) hereafter) were generally darker than at an earlier times (at 21:30), which could be attributed to some public and commercial light sources being turned off late at night. 


This survey not only provided the first glimpse of the light pollution situation in Hong Kong, but also spread the message of dark sky conservation and energy saving among students and the general public through participations in the hands-on sky brightness measurements and first-hand observations of the environmental consequences of light pollution. However, the dataset collected was limited by its geographic distribution (volunteers made measurements usually within or near urban population areas), temporal resolution (a majority of volunteers made measurements usually once or twice every several nights), a short monitoring time (volunteers were swapped every few months to allow for more participation), and possible human-related errors (volunteers might make mistakes during measurements and/or data reporting). 

We launched the succeeding project \textit{Hong Kong Night Sky Brightness Monitoring Network} (referred to as NSN hereafter) in 2010 to comprehensively study the properties of the NSB in Hong Kong and its dependence on time, location, and various atmospheric and meteorological conditions with the support of the Environment and Conservation Fund of the Hong Kong SAR government. The range, depth, and accuracy of data collection were optimized by setting up automatic NSB measurement stations in 18 distinct urban and rural locations around Hong Kong. All these stations were designed so that on-site NSB monitoring for over a year was possible after securing long-term commitments from our collaborating partners (refer to acknowledgment section for the full list). The temporal resolution of data collection was vastly improved and human errors were eliminated through the use of the ethernet version of the SQM. In Section~\ref{sec:methodology}, details of the survey are discussed, including the setup of the automatic NSB measuring stations, site selection, data selection, data quality control, and calibration of the measurement devices. The resulting database forms the backbone for a study of the overall light pollution conditions in Hong Kong. Results of the analyses can be found in Section~\ref{sec:results_analysis}, focusing on the contributions of artificial lighting sources on the observed NSB in an urban environment. Discussion of the results from this study and possible potentials of this NSB database are presented in Section~\ref{sec:discussions}. 

%% file: methodology.tex
\subsection{Data collection}\label{subsec:data_collection}

\subsubsection{Measurement instrumentation}\label{subsec:sqmle}

In the current study, the lensed version of the SQM, the Sky Quality Meter - Lens Ethernet (SQM-LE), was used to collect the night sky observation data. Both the SQM and SQM-LE measure the brightness of the night sky in units\footnote{\textit{Magnitude} (mag) is a logarithm scale international unit to measure the brightness of astronomical objects. A difference of 1 mag refers to an observed light flux ratio of 10$^{0.4}$ = 2.512. \textit{Arc second} (arcsec or ") is the unit of length on the celestial sphere. 1" = 1/3,600$^{\circ}$. Suppose the measured NSB at site A is 20.0 mag arcsec$^{-2}$, then the brightness of the sky is equivalent to a celestial object of 20 mag filling up a patch of sky of area 1 arcsec $\times$ arcsec. Suppose the measured NSB at site B is 19.0 mag arcsec$^{-2}$, then the sky at site B is 2.512 times brighter than that at site A.} of mag arcsec$^{-2}$. The SQM-LE is manufactured by the Canadian company \textit{Unihedron}\footnote{\url{http://unihedron.com/projects/sqm-le/}} and each unit was calibrated by the manufacturer before shipment. 
The light sensor of SQM-LE is the TAOS TSL237 High-Sensitivity Light-to-Frequency Converter
covered by a near-infrared blocking filter Hoya CM-500.
The combined filter-sensor system has a central wavelength of 540 nm and a Full Width Half Maximum (FWHM) of 240 nm \citep{sqmpdf}, and is designed to have a similar sensitivity to that of the human eye. Compared to the earlier version of the SQM, the SQM-LE has a narrower field of view of $\sim20^\circ$ in FWHM. The detector sensitivity drops sharply away from the zenith direction, with sensitivities at factors 15 and 100 respectively, lower than that at the zenith at 20$^\circ$ and 40$^\circ$ off-axis \citep{sqmlepdf}. The SQM-LE was preferred over the original unlensed version of the SQM for a NSB study in Hong Kong because the measurements would be less affected by effects of stray lighting directly reaching the light sensor in this dense urban environment. The stated accuracy of the SQM-LE by the manufacturer is 0.1~mag arcsec$^{-2}$, or roughly $\pm$10\% of the brightness of the sky in terms of emitted flux. 


While the convenience and the durability of SQM-LE favored their usage for a long-term night sky monitoring project, a few problems of this device had been identified. First, dropped transparency of the filter could occur due to frosting developed when operated in a humid environment for a prolonged period. A filter with anti-moisture coating had been provided by the manufacturer for replacement (cf Section~\ref{subsec:data_quality_calibration}). Second, the unit had not been calibrated to accurately measure a night sky darker than 23 mag arcsec$^{-2}$. However, this is a non-factor for our study as the night sky in Hong Kong would never reach that dark level. Finally, in earlier versions of the SQM-LE firmware (before version 19), the precision of the meter for readings brighter than $\sim$~17.0 mag arcsec$^{-2}$ suffered from digitization problems, causing the units to round off measurements between 15.2 and 17.0 mag arcsec$^{-2}$ to the nearest 0.15 mag arcsec$^{-2}$, and measurements brighter (smaller in NSB) than 15.2 mag arcsec$^{-2}$ to the nearest 0.18 mag arcsec$^{-2}$. As this problem had been fixed by a subsequent firmware upgrade, only $\sim~8\%$ of the total data collected at 13 different stations from 14 individual units in the first 20 months of the survey could be affected by this error. 

\begin{figure*}[ht]
\begin{center}
\includegraphics[width=14cm]{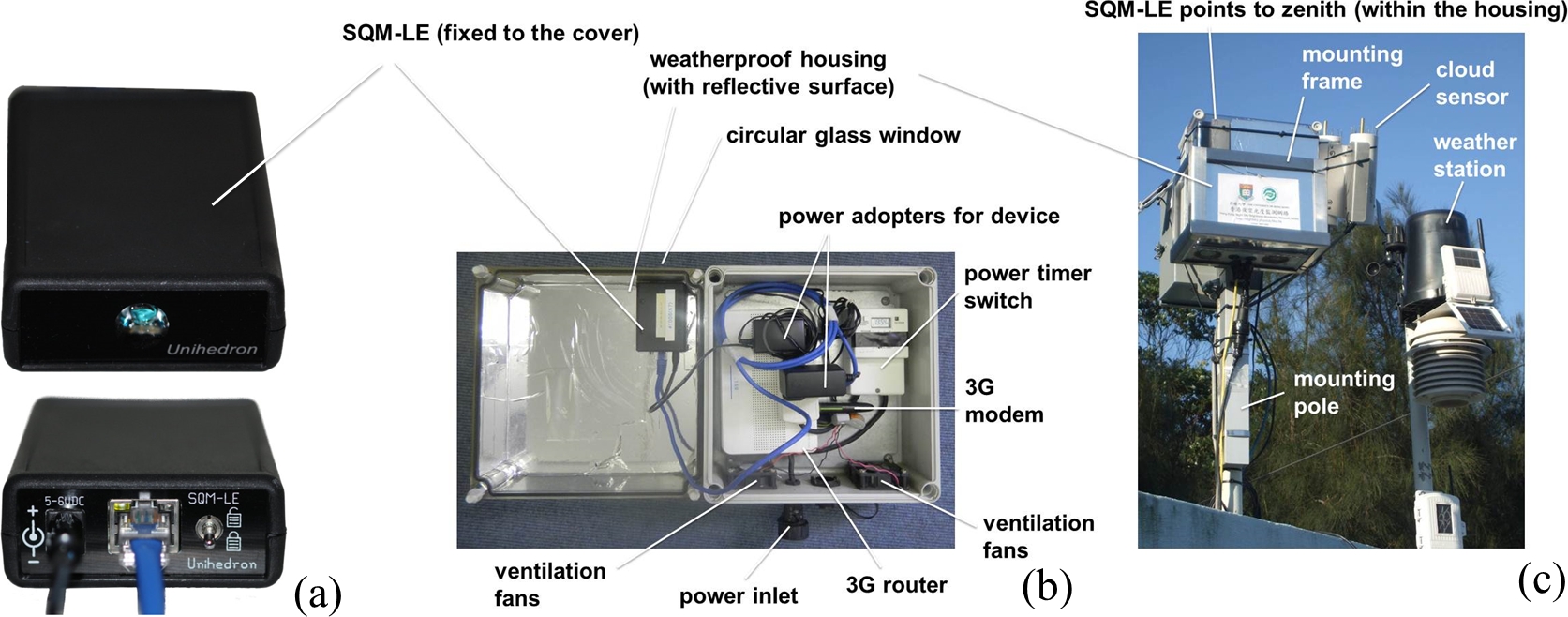}
\caption{A unit of SQM-LE is shown in \textit{(a)} (credit: \textit{Unihedron}). Its semi-conductor light sensor is located under the filter in greenish blue. The view in the bottom shows the power inlet, the ethernet inlet, and the switch for calibration purpose. The dimension of a SQM-LE unit is about 91 $\times$ 66 $\times$ 28 mm. The internal hardware configuration and appearance of the iObs NSN monitoring station are shown in \textit{(b)} and \textit{(c)} respectively. The dimension of this model of housing is 278 $\times$ 278 $\times$ 100 mm. As discussed in the main text, a small piece of the polycarbonate surface where the SQM-LE points towards the zenith had been replaced by a piece of a thin glass 25.4 mm in diameter. For each complete module, hardware costed about USD 770 in 2010, weights $\sim$~5 kg, and requires less than $\sim$~16 W AC power supply during normal operation. In the outdoor photograph, in addition to the night sky measurement module, two cloud sensors (narrow cylinders mounted on the right side of the rectangular housing) were also mounted on the left-side pole, while an automatic weather station installed by the Hong Kong Space Museum on the right-side pole was also shown. \label{fig:box_station}}
\end{center}
\end{figure*}

The SQM-LE was not originally designed for permanent outdoor installations. To protect it from the wear and tear of the outdoor environments, each unit was fitted into a waterproof polycarbonate housing
which has a transparent cover for light to reach the SQM-LE sensor. 
A unit of SQM-LE, the configuration and appearance of the setup were shown in Figure~\ref{fig:box_station}. The enclosure for each module also included a power supply adapter and a network device for data transmission (to be discussed in detail in Section~\ref{subsec:data_flow}). Except SQM-LE, almost all components of the observing module employed readily available commercial products to reduce the cost for construction and maintenance. The entire housing was supported on a tailor-made stainless steel frame mounted on a pole attached usually to the railings on the rooftop of buildings. The entire housing and frame could easily be dismounted so that the entire unit could be collected from an observing site to our laboratory for regular checking and maintenance. The modular design also allowed for swapping of hardware between stations.  

The summer temperature of Hong Kong can reach a high of 35$^{\circ}$C and thus temperature control is important for protection of the various electronic devices, including the SQM-LE (maximum operating temperature 85$^{\circ}$C). Even though the unit's temperature compensation algorithm is adequate under the expected outdoor operating conditions in Hong Kong \citep{schnitt:2013}, we installed ventilation fans at the bottom of the housing and glued silver reflective linings to reduce the heat accumulation in the interior of the unit. We placed the linings inside the housing instead of outside to ensure that they would not be detached after long-term outdoor exposure, and it had proved to serve its purpose. In each module, an electronic timer switch limited the power supply to 15:50 $-$ 09:00 each day for both energy saving and for reducing heat generation. The daily power reset also occasionally helped to overcome system failures during the night, and recovered operation of the module in the next day to reduce the number of required manned on-site service maintenance. 

Through regular maintenance of the observing modules collected from the monitoring stations, it was discovered that the transparent cover window of the polycarbonate housing suffered from obvious aging and turned visually yellow after $\sim$~12 months of outdoor operations. To prevent further degradation in data quality, a small portion of the polycarbonate housing right on top of SQM-LE was replaced with a 25.4 mm in diameter and 3 mm thick high efficiency circular glass window with anti-reflection coating (Edmund Optics Stock No. \#46-098\footnote{\url{http://www.edmundoptics.com/}}).  The glass was glued directly above the SQM-LE sensor by drilling a small hole on the housing cover. Procedures for corrections of the data that suffered from the aging problem will be described in detail in  Section~\ref{subsec:data_quality_calibration}. 



\subsubsection{Monitoring stations}\label{subsec:obssite}

From our previous study of the light pollution condition in Hong Kong~\citep{pun:2012}, it was clear that the level of external lighting was the dominating factor of the NSB level observed. With a highly mixed land utilization within a small landmass, along with a complex landscape and diverse forms of human activities in different areas in Hong Kong, it was desirable to strategically choose monitoring stations to form a network covering as wide a geographical coverage, range of population density, landscape type, and land use as possible within the budget constraint. 

Apart from the demographic and geographic considerations described above, each observing module should have a wide field-of-view (FOV) to the sky not obstructed by any lighting source. This could pose a big challenge for our urban monitoring site selection due to the high density of buildings in the city residential areas. In almost all cases the measurement module was placed on the rooftop of a building for the widest FOV. Moreover, the ambient environment near each monitoring station was checked so that the top floor of the nearest building was at least 40$^\circ$ away from zenith. As discussed in Section~\ref{subsec:sqmle}, the sensitivity of the SQM-LE at that angle was only a factor of 0.01 of that at the zenith. All monitoring stations satisfy this criteria, except in the WTS where the rooftops of a handful of buildings were visible at an zenith angle of $\sim$~30$^\circ$. Other factors for consideration included the availability of power, and more importantly, steady and strong mobile Internet network at the observing locations (cf Section~\ref{subsec:data_flow}). These could be demanding requirements for some of the rural locations in our survey. 


With a hilly and complex terrain, the population density of Hong Kong is highly uneven. Regions of the highest population density lie on either side of the Victoria Harbor between the Hong Kong Island and Kowloon. Additional population centers spread throughout the rest of the city (known as the New Territories region), observable as clusters of light in the ISS image in Figure~\ref{fig:station-map}. On the other hand, large portions of the territory are allocated as reserved park areas and remain sparsely populated. In an ideal situation, the monitoring network should cover as wide a geographical distribution and population density as possible. However, with limitations of manpower, funding, and accessibility to suitable sites, we ended up selecting 18 locations for construction of NSN stations. Details of these stations were listed in Table~\ref{tab:station-list}. In Figure~\ref{fig:station-map}, the station locations were overlaid on the night-time image taken from ISS in 2003. Station codes in Table~\ref{tab:station-list} will be used to refer to specific stations hereafter. While this project was conducted between 2010 and 2013, almost a decade after the ISS image in Figure~\ref{fig:station-map} was taken, it is expected that the overall pattern or distribution of population, henceforth the amount of external lighting used, did not change by much for a majority of the regions during the intervening period. 

Unlike many cities worldwide, the wide spread of population centers in Hong Kong made the exact classification of each monitoring locations as urban, suburban, and rural to be ambiguous. On the other hand, the ISS image did seem to justify the classification of only two categories, urban and rural.   
The classifications were listed in Table~\ref{tab:station-list} and shown in Figure~\ref{fig:station-map}. With this rough system, a total of 10 monitoring stations were classified as urban and 6 as rural. In two instances, the station was neither classified as urban nor rural because of their unique locations near the Hong Kong International Airport (GFS) and the River Trade Terminal (TMD). 



Brief descriptions of all the monitoring stations listed in Table~\ref{tab:station-list}, in their order of operation start dates, are provided below.

\begin{description}

\item[KP] This urban station was located at the King's Park Meteorological Station of the Hong Kong Observatory (HKO). Situated on the summit of a small hill at the middle of Kowloon, it offers a location with a relatively wide FOV in the middle of a densely populated mixed residential and commercial area. A large variety of surface meteorological measurements are conducted daily by the many equipment operated by the HKO at this station\footnote{\url{http://www.hko.gov.hk/wxinfo/aws/kpinfo.htm}}.


\item[TC] This urban station was located in a satellite residential town adjacent to the Hong Kong International Airport. The NSN module was installed at the rooftop of a two-storey government building in the center of the town. In addition to shopping malls and high-rise residential buildings nearby, the major artificial light source is the Airport passenger terminal at about 3~km away. 

\item[HKn] This urban station was located near the astronomical observatory on the rooftop of an environmental education center complex on the foothill of Tai Mo Shan, the highest peak in Hong Kong. While it has the highest elevation ($\sim$~140~m above sea level) among all stations in the network, it was still affected by external lighting of even taller highrises nearby and from the adjacent town center. The education center held regular school group and public visits which could occasionally affect the data taken. 


\item[HKU] This urban station was installed near the teaching astronomical observatory on the rooftop of the Department of Physics building at the University of Hong Kong (HKU). HKU has an urban campus with high-rise residential buildings in the surrounding. Meteorological equipment including two cloud sensors, an all sky camera, and an automatic weather station were installed near the night sky monitoring module for additional measurements of the atmospheric conditions. 
Occasional educational events were held at the observatory which could affect the operation of the monitoring station. 

\item[iObs] This rural station was installed on the rooftop of the building housing the robotic observatory of the Hong Kong Space Museum inside a rural country park. The observatory is part of a complex of buildings in a public camping village. The main sources of artificial lighting include the external fixtures of the camp cottages and recreational facilities within the complex. Moreover, scattered lights from a town centre about 6 km away in southwest (Sai Kung) can be noticed from the observatory. Similar to the HKU station, meteorological equipment including two cloud sensors, an all sky camera, and an automatic weather station were installed near the night sky monitoring module (the latter two equipment were installed by the Hong Kong Space Museum). As part of a public educational facility, stargazing events were regularly held as part of the Museum and village programs.  


\item[TST] This urban station was installed on the rooftop of the Hong Kong Space Museum. The museum is located in the waterfront of the Victoria Harbor, which is the central commercial area of the city clustered with a large number of hotels, shopping malls, and tourist attractions. 
The Museum building is relatively low in height ($\sim$10 m) compared to office buildings and hotel towers nearby that can reach $\sim$100 m. In addition to the spilled light from the interiors of these buildings, other lighting sources include the large and bright illuminated billboards decorating the building exterior, particularly near the rooftop of the buildings. Similar to the iObs station, stargazing and educational activities at the rooftop could occasionally affect the measurements. 

 
\item[SH] This rural station was located on the rooftop of the building housing the observatory of the Hong Kong Astronomical Society, which was operated by local amateur astronomy enthusiasts in a rural village setting, with only a small number of houses in the vicinity. 
While the Airport is located about 10~km away to the North, this site is not expected to be strongly affected by its strong lighting due to the shielding effect from the Lantau Peak (altitude 934 m) in between. Members of the Society might use the observatory for observations during clear weather nights and these operations could briefly affect the data collected.

\item[Cap] This rural station was located on the rooftop of the marine science research center operated by HKU at one of the southeastern tips of the Hong Kong Island. The building lies on the shore of a marine reserve area with minimal amount of artificial lighting fixtures in the ambient environment to allow for ecological studies of the region. The main light source is a lighthouse (at a height 20~m above the observing module) located at $\sim$~200~m away.


\item[AP] This rural station was located in the Astropark, a designated stargazing park designed by the Hong Kong Space Museum. The AP station was located $\sim$~3 km deeper into the same country park compared to the iObs station described earlier. No external lighting operates within the park, but scattered light from a water sports center at 100~m away could affect measurements taken at the station. 
Since its opening in 2010, the park has been one of the most popular stargazing venues in Hong Kong for astronomy enthusiasts.

\item[TMD] This station was located in a temporary observatory setup near the Hong Kong Science Museum's depot in western New Territories. While the nearest residential town was relatively distant at $\sim$~4~km away, the station's lighting environment was mainly affected by the River Trade Terminal at about 0.5~km away. This container terminal for river barges has 49 berths and a large number of lighting fixtures remained turned on throughout the night. Therefore this station is classified as neither an urban nor a rural site.


\item[MWo] This rural station was installed on the rooftop of a campsite at about 1 km away from a ferry pier, which was also the location of a rural town center. This station was closed permanently in August 2011 due to the closure of the campsite for major renovation work.

\item[TPo] This urban station was installed on the rooftop of a government office building located within a population center in the New Territories. Public facilities and residential estates in form of high-rise apartments as close as 30~m away could affect the measurements collected.


\item[TSW] This urban location was established within the Hong Kong Wetland Park which is a conservation, education, and eco-tourism facility. The park comprises of a visitor center and 60 hectares of outdoor wetland protected areas. The station was installed in a location inside the park that was not open to the pubic. 
As an ecologically important area, usage of outdoor lighting within the park is kept to a minimum. On the other hand, a major population center is located about 500~m away from our detector and could contribute to the observed NSB. 

\item[WTS] This urban station was installed on the rooftop of a school located in a densely populated district with residential estates, shopping malls, and transportation hubs in Kowloon. 
At only nine storeys high, our detector was practically surrounded by taller buildings in all directions, with the closest one at about 70~m away extending up to $\sim$~30$^\circ$ in zenith angle.

\item[TpM] This rural station was installed on the rooftop of a Police Post on a sparsely populated (population $\sim$~100) island on the northeastern corner of Hong Kong. The entire island was inside a country park, with limited artificial lighting sources apart from public facilities. The remoteness of the site also led to our selection of a different data acquisition mode for this station due to the limited mobile Internet strength (cf Section~\ref{subsec:data_flow}).

\item[SSh] This urban station was installed on the rooftop of a school located in a population center in northern New Territories. Situated near the edge of town center, this observing location has relatively wide FOV. This station was the northernmost in our study, and its data could also be affected from the city of Shenzhen, China, at only 4~km away.

\item[ST] This urban station was installed on the rooftop of a  school located in a population center in southern New Territories. This school is also surrounded by taller residential buildings, with the closest block at about 80~m away. 

\item[GFS] This station was installed within the confines of the Hong Kong International Airport. This busy airport operates 24 hours a day and most of the external lighting for air traffic remains turned on throughout the night. This station was installed on the rooftop of a facility building on the southwestern corner of the airport. The South runway and the airport passenger terminal are located about 500 m in North and 3 km in northeast. 
Both the HKO and the airport operator (Airport Authority Hong Kong) installed suites of advanced meteorological and air quality monitoring systems at different locations of the airport for air traffic operations and for environmental monitoring. We also installed cloud sensors adjacent to the SQM-LE for further studies. 

\end{description}

\input{table-station-list}



The lighting environment near each monitoring station were recorded at the beginning of the study and monitored regularly through our maintenance trips to ensure that the FOV of all the SQM-LE were not blocked by any obstacles such as trees or artificial structures. Moreover, we observed no evidence of drastic change of the lighting settings in our monitoring stations throughout up to 3 years of duration of survey. 



\subsubsection{Data flow}\label{subsec:data_flow}

The NSB measurements taken from the SQM-LE could be transferred by Internet directly via a TCP connection to the survey computer server (the main server). With a majority of monitoring stations located at third-party properties, the mobile 3G network was used for Internet connections to all the monitoring stations to minimize the cost and labor required. 


For all but one of the monitoring stations, we adopted the \textit{active} mode of data transfer. Each SQM-LE unit acted as a TCP client, linking first to a commercial 3G router, and then a USB 3G modem for Internet connection. Each unit would then be assigned a static IP address. The main server was programmed to request measurement from all NSN stations simultaneously every day from 16:00 to 08:59 in the next day at regular time intervals. The start and end time of operation were intentionally set to be earlier than the sunset time and later than the sunrise time respectively throughout the entire year in Hong Kong. The data collection frequency was initially set to be once per 1 to 5~minutes during the early testing stage, and was unified to once every minute since late 2011. 

The SQM-LE would respond to the request from the main server by sending a formatted string of results, which included the NSB reading, sensor temperature, device serial number, and other information back to the main server. A new entry was created in a \emph{MySQL} database for each returned string. The \textit{timestamp} of an individual entry was defined to be the time when the main server received the returned string. Normally the string arrived in less than 10 seconds. Four consecutive data requests for a specific station would be made by the main server if it did not receive the data string in its prior attempt(s). An error log would be generated if all attempts failed before waiting for the start of the next data requesting cycle. 


In the remote rural monitor station TpM, a \textit{passive} mode of data transfer was adopted due to the weak mobile network coverage for the Internet service provider that serviced all the other stations.
While the hardware configuration remained the same, the data flow out of TpM was different. The networking enhancement featured in the SQM-LE firmware version $\geq$ 13 was adopted so that the device was able to act as a TCP server and send measurement readings to the main server via Internet at regular time intervals. 
To collect data from the TpM station, a \textit{listener} program ran continuously on the main server to receive data strings coming in from it. The data sampling interval at this station was set as once per 5 minutes because of the limited Internet bandwidth. 
The device could still make measurements even in the case of disrupted Internet connection by storing the unsent data strings in an internal memory . Once the Internet connection resumed, all the stored measurements would then be sent out simultaneously. 
An adjustment program was developed to fix the \textit{timestamp} of these data strings for storage in the \textit{MySQL} database.

\subsection{Data selection}\label{subsec:data_select}

The stability and reliability of NSB measurements were crucial for establishing a long-term database of NSB in Hong Kong. Special attention was paid to maximize the number of measurements collected while ensuring the quality of data received. In this section, we describe our efforts to monitor and improve the success rate of data collection. Moreover, we discuss how we filtered out entries in our database that were due to \textit{non-night-sky} entries taken under the influence of sunlight, or during the periods of known non-routine human activities, testing, or hardware failures, etc., before any meaningful analysis could be performed. 

As discussed in Section~\ref{subsec:data_flow}, all \textit{active} stations operated between 16:00 and 08:59 (17 hours) daily during the production stage of the NSN. For a data collection frequency of once per minute, a total of $17 \times 60 = 1,020$ individual data entries would be expected from each station for each night. In reality, data taking failures could be caused by a multitude of factors, including faulty SQM-LE units and/or mobile network hardware, power supply failure, network failure, or any combinations of them. The data collection success rate for a particular night of each station was defined to be the number of entries received divided by the expected number (1,020). By daily monitoring of this value for all stations, we were able to identify problematic stations and take appropriate actions to resume data collection as soon as possible. 


After settling on a 1 minute data-taking frequency in December 2011, the monthly success rate of stations dropped to values below 50\% (i.e., less than half of expected number of entries received) for only roughly 5\% of the time. While it was difficult to accurately diagnose, the exact reasons behind this drop which was usually discovered the day after. In most of the cases, these stations would resume normal operation and reconnect to the network after the scheduled system reboot by the timer switch in the system the next afternoon (cf Section~\ref{subsec:sqmle}). On rare occasions of a persistent drop in the success rates, site visits would be arranged for on-site servicing or for replacement with a back-up complete module if the problem could not be solved in-situ. Through replacement tests with a backup SQM-LE, we found that the origin of the problem was usually due to unstable mobile network connection at a particular location or an unsteady network router/USB-modem. In summary, we attempted to maximize the amount of data taken given the limitations of manpower and the structure of the data acquisition system, and achieved a success rate of around 86\% for the network near the end of the study. 


With each NSN station operating before sunset and after sunrise daily, data entries collected under the influence of sunlight had to be excluded. The sky would be totally free from sunlight between astronomical dusk and astronomical dawn, that is, when the altitude of the center of the Sun is $\geq 18^\circ$ below horizon, which translates to  
74 to 86 minutes after sunset and before sunrise throughout the year in Hong Kong \citep{hkotwilight2011}. However, based on actual measurements of the sky brightness taken near summer and winter solstices, it was concluded the conditions could be slightly relaxed to 70 minutes after sunset and 70 minutes before sunrise, due to obstructed near-horizon sightlines reaching the light sensors. While these sunlight-affected entries, making up 40\% of all collected, would be excluded for NSB studies, they could potentially be used in the future for studies of atmospheric optical properties through changes in the sky brightness level during twilight~\citep{patat:2006,rozenberg:1966,hulburt:1953}.

Apart from sunlight, data entries that were classified as taken during \textit{non-night-sky} conditions are excluded. These included entries affected by non-routine human activities. As described earlier, some of the monitoring stations were installed near astronomical facilities where public or private observing sessions were conducted. The duration of these activities was usually short (several hours) and ended before midnight. Data entries taken during these activities were all removed even though the use of lighting was usually kept at a minimal level during an observing session. On the other hand, emission from celestial objects could also contribute to the background NSB. The Moon is the major contributor to the NSB. In our database, data from all Moon conditions, including altitude, phase, or even for cases of lunar eclipses (there were total of 6 lunar eclipses, including a penumbral eclipse, visible in Hong Kong during the project period \citep{nasalunareclipse}) would be included (the  effects of moonlight on NSB distribution will be discussed in Section~\ref{subsec:moon_nsb_distribution}). 
Finally, there were occasional abnormal NSB ($>22$ or $<10$ mag arcsec$^{-2}$) entries collected even after the exclusion of sunlight affected entries. The exact origins of these unphysical readings are unknown, though likely to be electronic glitches of the SQM-LE or the data acquisition hardware. On the other hand, other explanations such as lightning or obstruction by nocturnal animals or insects could not be excluded.  


In summary, only about 5\% of all sunlight-free entries collected were considered as \textit{non-night-sky} events. We believe that entries taken during any unreported or missed \textit{non-night-sky} events (for example, unreported stargazing sessions) only accounted for a minute subset of the entire database. We can therefore conclude that statistically speaking, our night sky database truly reflects the variations of the NSB due to a large range of artificial conditions (use of external lighting in public and private facilities) and natural phenomena (for example, variations in moonlight, cloud amount, or rain).




\subsection{Data quality and calibration}\label{subsec:data_quality_calibration}

\begin{figure*}[ht]
\begin{center}
\includegraphics[width=14cm]{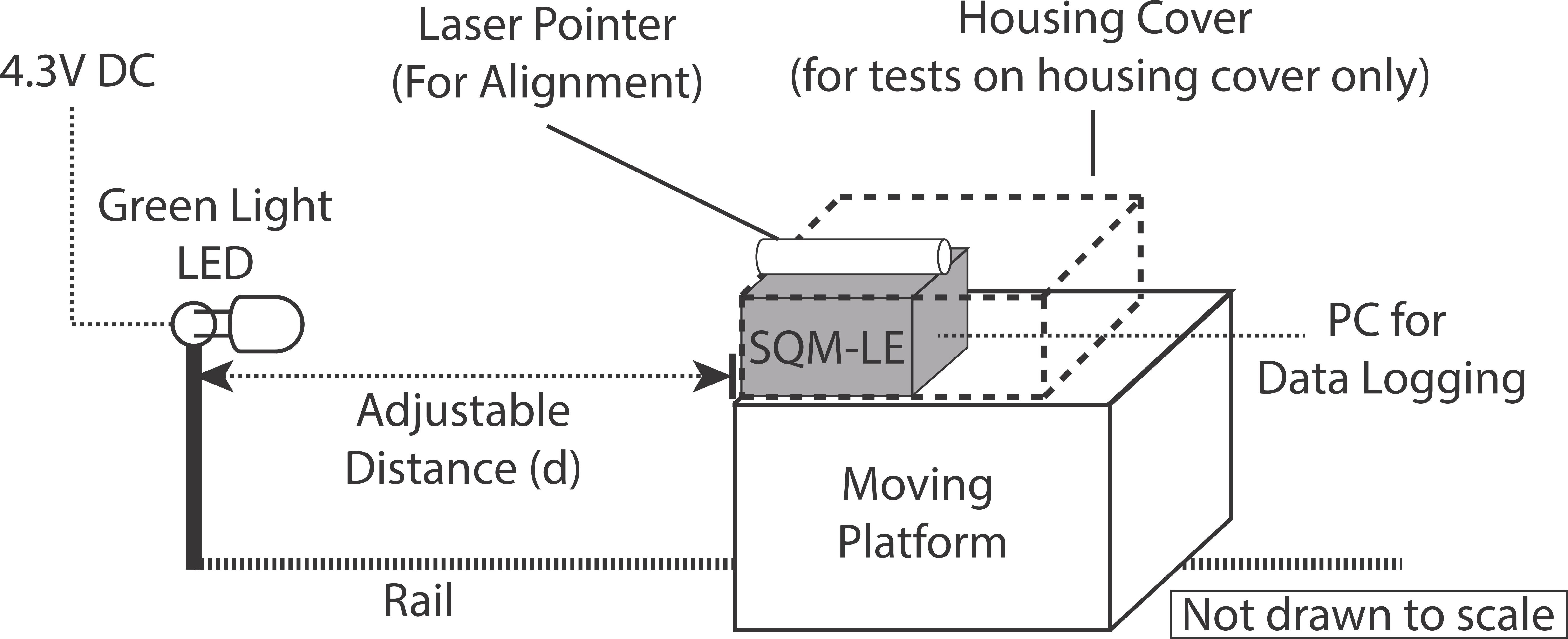}
\caption{Schematic drawing of the calibration and testing platform for the SQM-LE and the covering optical components. The light sensor (and housing cover) will be placed on the moving platform during experiment. The whole setup was installed within a metal frame and totally light-shielded by double layer of dark cloths (not shown here). The metal frame had a size about 1.1 (H) $\times$ 1.0 (W) $\times$ 2.1 (D) m. The moving platform could be moved on the rail so that the distance between the light sensor and the LED light source $d$ can be modified. Refer to the main text in Section~\ref{subsec:data_quality_calibration} for details on calibration and testing. \label{fig:test_stand}}
\end{center}
\end{figure*}

Designed to be a long-term survey, it was extremely important to ensure the quality of the data received for long-term comparison studies. All components of the NSN monitoring stations had to be exposed to an outdoor environment for at least a year. One major challenge was to monitor the effects of sunlight exposure on the optical properties of the various components, from the near-infrared blocking filter of SQM-LE unit to the housing window through which the SQM-LE took the NSB readings. To measure the change in transparency of the optical components over time, we developed a calibration and testing platform in our laboratory. The platform comprised of a standard green LED light source (central wavelength 532 nm, FWHM 31 nm) and a moving platform, with the set-up shown in Figure~\ref{fig:test_stand}. The distance between the light source and the SQM-LE sensor, $d$, could be adjusted by rails beneath the moving platform, on which the SQM-LE (together with the housing cover if we are conducting measurements on the cover) could be anchored. The alignment of the optical components was checked by laser pointers. The entire platform was covered by double-layer of thick black cloth during testing for light-shielding and to minimize internal light reflection within the setup. 

It should be noted that the light source used for our testing platform was a point source (with light intensity dropping off from the center with FWHM 40$^\circ$) rather than a flat diffused source lit by fluorescent light or an 8-inch uniform integrating sphere used in the experimental setups of the manufacturer of the SQM-LE (Anthony Tekatch 2009, personal communication) and in \citet{sqmpdf} respectively. This could pose slightly bigger errors in case of misalignment between the light source and the detector. On the other hand, the range of SQM-LE readings checked in our setup at $\sim$~16 to 18 mag arcsec$^{-2}$ would be closer to the actual NSB observed than the range studied by the calibration setups of the manufacturer and in \citet{sqmlepdf}. Moreover, the spectral characteristic of our light source was very different from that of the typical light-polluted night skies \citep{sheen:2004,slanger:2003,della:1999,lane:1978}. Nevertheless, performance of all optical components and SQL-LE units used in our study were monitored under the current design since the start of the project. 

All SQM-LE units had been calibrated by the manufacturer before arrival, with a claimed absolute precision of NSB measurement at $\pm$ 0.10 mag arcsec$^{-2}$, or roughly $\pm$10\% in terms of light intensity. The SQM-LE units, together with the polycarbonate housing, were collected from all the monitoring stations to the laboratory testing platform at a frequency of roughly every $6 - 9$ months for servicing. The collected SQM-LEs were placed at various distances $d$ from the light source without any cover, making 15 consecutive measurements at each $d$ to yield the average reading at that distance. 
Average measurements from different units were compared to look for any abnormality. At one of the servicing trips, it was found that the transparency of the filters of the first two SQM-LE units installed reduced drastically (up to 1 mag arcsec$^{-2}$) due to frosted filter covering the light sensor. It was believed to be due to tiny particulates in the atmosphere that were carried to the filter glass by atmospheric moisture (Anthony Tekatch 2011, personal communication). These units were shipped back to the manufacturer with the affected filters replaced by anti-moisture coated versions with minimal disturbance to data collection. Unfortunately the exact patterns of the drop in performance of these two sensors were not known and could not be traced. On the other hand, all data taken by these two meters during the possibly affected period (73,000 for the KP and 72,000 for the TC station, making up less than 1/4 of the total collected, cf Table~\ref{tab:statistic-results}) were studied and it was found that the overall average NSB of the two stations did not change even if these entries were removed. Therefore we decided to retain these data in the database and did not distinguish them in the subsequent analyses. All other units subsequently shipped were already pre-installed with anti-moisture filters. All the SQM-LE units were shipped back to the manufacturer for a final calibration after the end of the project and the results were consistent in general with the initial accuracy claims of the manufacturer.



For accurate readings of the NSB, the small decrease in light intensity due to the application of the cover have to be corrected. A similar testing procedure was applied to study the transparency of the polycarbonate cover windows of the monitoring housing. 
The amount of light attenuation due to the cover was deduced by comparing measurements made with and without the cover at different distances $d$ to the light source. 
It was found that the transparency of the polycarbonate material used for the water-tight enclosure cover dropped over time after several months of exposure, as easily witnessed from the color of the windows turning yellow. This was most likely due to interactions of the materials with UV radiation from sunlight after prolonged exposure. 



\begin{figure*}[ht]
\begin{center}
\includegraphics[width=14cm]{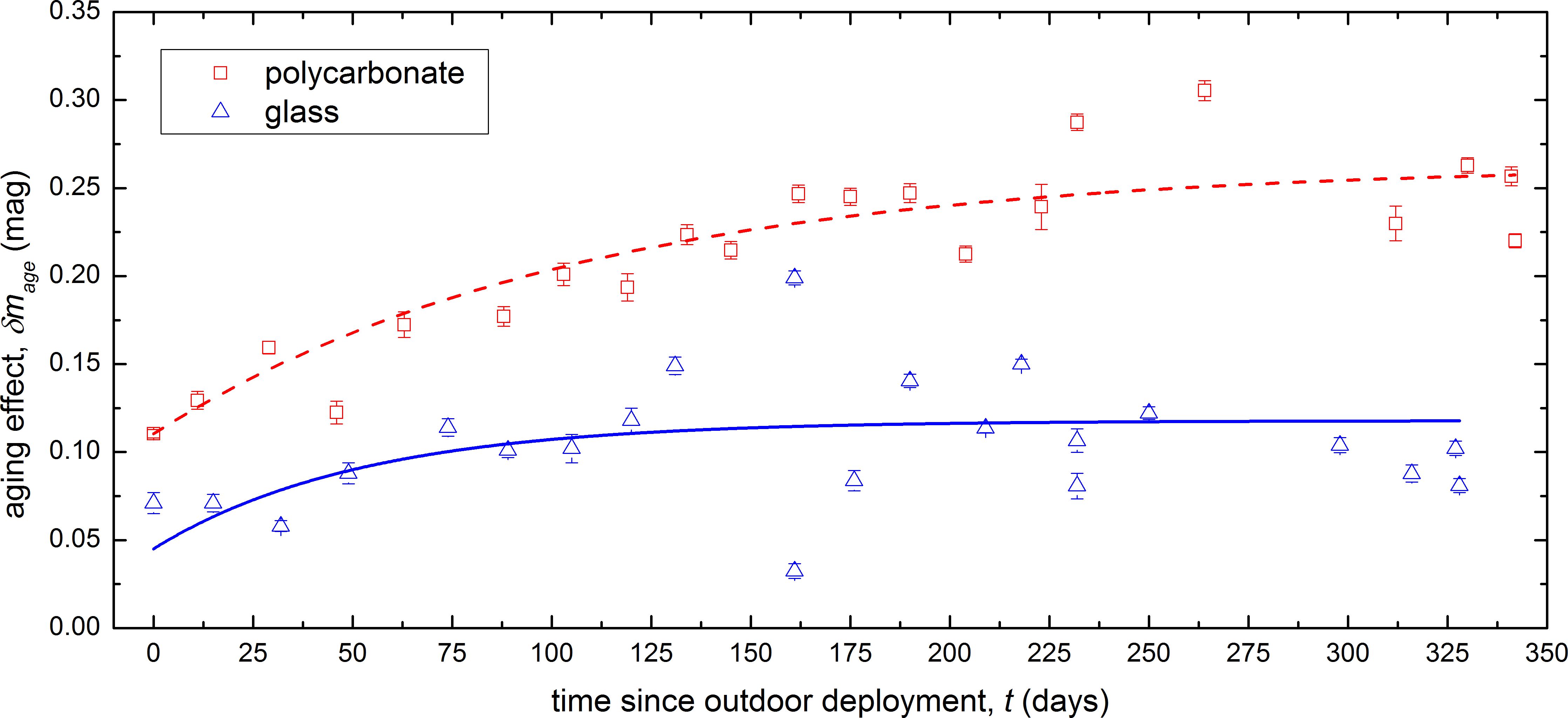}
\caption{Repeated measurements on the light attenuation of the same housings with polycarbonate cover (\textit{squares}) and glass window (\textit{triangles}) reveal the their agings over time due to prolonged exposure of UV radiation from the Sun. Their best-fit model curves for the time-dependent attenuation (Equation~\ref{eqt:aging}) were overlaid (\textit{dashed} and \textit{solid} curves respectively) for comparison. The error bars described the spread of data points during experiments,  which were conduced after surface cleaning.  \label{fig:attenuation}} 
\end{center}
\end{figure*}

To correct for this problem, two remedies were taken: first, all the polycarbonate covers were replaced with a 1-inch diameter high throughput glass windows starting from February 2012; second, new units of the polycarbonate housing with and without glass window were installed outdoor so that changes in its optical properties over time could be studied. Results of these tests were summarized in Figure~\ref{fig:attenuation}. Even for a newly deployed unit, the polycarbonate cover carried a larger dimming effect $\delta m_{\rm age}(t = 0) \sim$~0.10 mag arcsec$^{-2}$ compared to the glass window (0.05 mag arcsec$^{-2}$). Concluding from over 300 days of outdoor deployment and measurements, the glass window did suffer from less aging compared to the polycarbonate cover. It was determined that the long-term optical attenuation for the polycarbonate covers and the glass windows leveled at $\sim$ 0.25 and 0.10 mag~arcsec$^{-2}$ respectively, nearly 1 year after the commencement of sunlight exposure.

The aging effects ($\delta m_{\rm age}$) of the polycarbonate cover and the replaced glass windows could be modeled as functions of the number of days of sunlight exposure ($t$) in the form 
\begin{equation}
	\delta m_{\rm age}(t) = A - B \exp[C(t-t_o)] \label{eqt:aging} \, ,
\end{equation}
where the parameters $A$, $B$, $C$, and $t_o$ were determined by nonlinear least squares fitting. Their best-fit values were tabulated in Table~\ref{tab:aging_fit}, while the best-fit curves were overlaid on Figure~\ref{fig:attenuation} for comparison. All measurements in our database have been corrected for the dimming effects of these components, through careful calculations of the number of days the particular material, be it polycarbonate or glass, spent in the outdoor environment by the same Equation~\ref{eqt:aging}. This ensured the integrity of the entire database for studies of long-term variations of the NSB.

\begin{table}[htbp]
	\centering
	\caption{Best-fit parameters for Equation~\ref{eqt:aging} which model the aging of housing optics as a function of exposure days $t$.}
		\begin{tabular}{c c c}
			Parameter & Polycarbonate cover & Glass window \\
			\hline
			$A$ & 0.26 & 0.12 \\
			$B$ & 0.40 & 0.03 \\
			$C$ & -0.01 & -0.02 \\
			$t_o$ & -102.34 & 48.67 \\
		\end{tabular}	
	\label{tab:aging_fit}
\end{table}

We also investigated temporal variations of the NSB readings observed due to different amount of dust deposited on the surface of the protective casing window. This was an issue since almost all monitoring stations were at unmanned sites and could only be cleaned during our servicing and by natural processes such as rain. We compared the difference in the attenuation by the casing window (either polycarbonate or glass) right after the stations were collected for servicing and after they were washed and cleaned. It was found that the effects were generally small at less than 0.1 mag arcsec$^{-2}$.

%% file: table-station-list.tex
\begin{sidewaystable*}

\caption{Geographical details of the NSN monitoring stations.} 
\centering 
\begin{tabular}{c c c c c c c c}
\hline\hline 

Code 	&	Facility	&	 Latitude N 	&	 Longitude E 	&	 Elevation	&	Operation Date	\\
	&		&		&		&	(m above sea level)	&	(DD-MM-YY)	\\
\hline 											
 \multicolumn{6}{c}{\textbf{Urban}}	 \\				
KP   	&	King's Park Meteorological Station, Hong Kong Observatory	&	22.3120	&	114.1723	&	65	&	26-05-10	\\
TC   	&	Tung Chung Monitoring Station, Environmental Protection Department	&	22.2888	&	113.9435	&	30	&	11-08-10	\\
HKn  	&	Ho Koon Nature Education cum Astronomical Centre	&	22.3836	&	114.1080	&	140	&	28-08-10	\\					
HKU  	&	Observatory Dome, HKU	&	22.2833	&	114.1398	&	100	&	01-09-10	\\
TST  	&	Hong Kong Space Museum	&	22.2942	&	114.1715	&	10	&	15-09-10	\\
TPo 	&	Tai Po Monitoring Station, Environmental Protection Department	&	22.4510	&	114.1646	&	25	&	10-12-10	\\
TSW 	&	Hong Kong Wetland Park &	22.4669	&	114.0087	&	5	&	21-12-10	\\
WTS  	&	Our Lady's College 	&	22.3452	&	114.1965	&	50	&	15-03-11	\\
SSh  	&	Elegantia College (Sponsored By Education Convergence)	&	22.4925	&	114.1242	&	10	&	22-03-11	\\
ST  	&	POH Chan Kai Memorial College	&	22.3664	&	114.1781	&	20	&	30-03-11	\\
  		&		&		&		&		&	 	\\											
\hline 											
 \multicolumn{6}{c}{\textbf{Rural}}	 \\												
iObs   	&	iObservatory, Hong Kong Space Museum	&	22.4083	&	114.3229	&	80	&	02-09-10	\\
SH  	&	Shui Hau Observatory, Hong Kong Astronomical Society 	&	22.2221	&	113.9153	&	15	&	27-09-10\\
Cap  	&	The Swire Institute of Marine, HKU	&	22.2079	&	114.2603	&	5	&	11-11-10	\\
AP  	&	Astropark, Hong Kong Space Museum	&	22.3766	&	114.3361	&	5	& 12-11-10	\\
MWo 	&	HKPA Silvermine Bay Camp	&	22.2736	&	114.0031	&	25	&	26-11-10	\\
TpM  	&	Tap Mun Monitoring Station, Environmental Protection Department	&	22.4713	&	114.3607	&	15	&	22-03-11	\\
  		&		&		&		&		&	 	\\			
\hline 											
 \multicolumn{6}{c}{\textbf{Not classified}}	 \\		   											
TMD  	&	Tuen Mun Government Depot, Hong Kong Science Museum	&	22.3679	&	113.9432	&	15	& 16-11-10
	\\
GFS    	&	Government Flying Service, Hong Kong International Airport	&	22.2961	&	113.9101	&	10	&	31-05-11	\\  
			
\hline 
\end{tabular}
\label{tab:station-list} 

\end{sidewaystable*}

%% file: results-analysis.tex
\subsection{Overall results}\label{subsec:overall_results}

\begin{figure}[ht]
\begin{center}
\includegraphics[width=8cm]{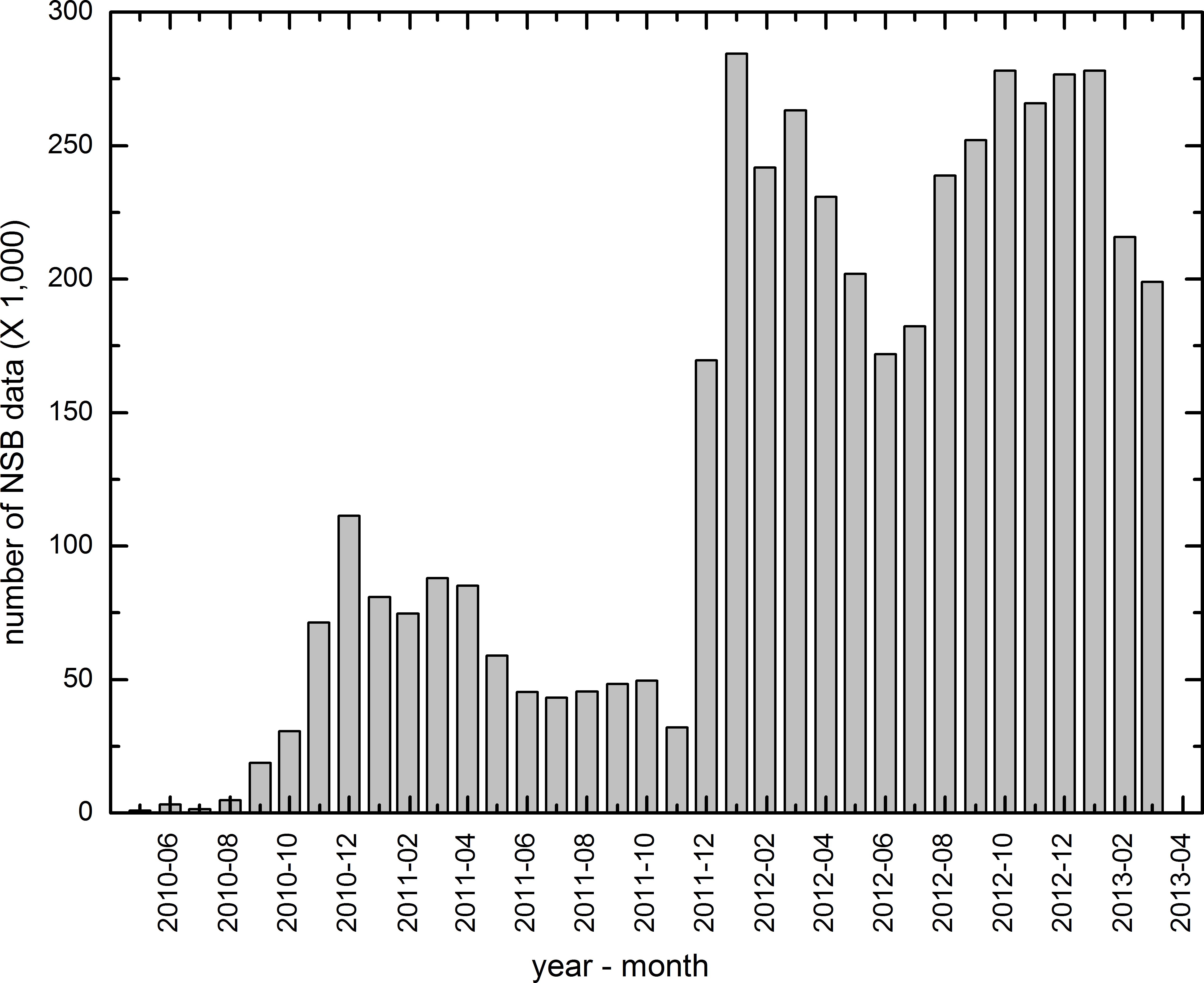}
\caption{Histogram of the monthly NSB data collected from all the monitoring stations. The data taking frequency was increased from once every 1 to 5 minutes to once every 1 minute for all stations since mid-December 2011, resulting in the sharp increase in the number of data points collected afterwards.
\label{fig:monthly_stat}}
\end{center}
\end{figure}

A total of almost 8.6 million raw data entries were collected from the 18 monitoring stations since operation of the first station at KP in 26 May 2010 to the end of the study in 31 March 2013. Of the data entries received, 4.6 million (54\%) were recording real NSB that were not affected by the Sun or by human interference (cf Section~\ref{subsec:data_select}). The monthly accumulation of the NSB data throughout the duration of the project is shown in Figure~\ref{fig:monthly_stat}. As presented in Table~\ref{tab:station-list}, the commissioning of the monitoring stations was conducted in phases, with a majority of stations in operation by late 2010. In the first few months of the survey, various data taking frequencies were tested until settling down to once every 5 minutes for a total of $\sim$~40k~-- 50k data points per month. As described in Section~\ref{subsec:data_flow}, the data collection frequency of the entire network was raised to once per minute starting from late 2011, hence increasing the number of NSB received to $\sim$~200k~-- 250k data points per month.  The monthly variation of the data sample collected was mostly due to the downtime of a few of the monitoring stations caused by various hardware or network-connection issues. 

Of the NSB data collected, about 3.0 million (64\% out of the real NSB data received) came from the 10 urban station, 1.1 million (24\%) from the 6 rural stations, and 0.5 million (12\%) from the two stations that were neither classified as urban nor rural due to their peculiar locations. The numbers of NSB measurements from each station were presented in Table~\ref{tab:statistic-results}. Similar numbers of NSB data were received from most of the stations except two: the MWo station due to its early termination (cf Section~\ref{subsec:obssite}), and the TpM station due to the lower data collection rate for that particular station (cf Section~\ref{subsec:data_flow}).

\input{table-statistic}

The average and standard deviation of the NSB observed in each monitoring station were also tabulated in Table~\ref{tab:statistic-results}. The overall average NSB was 16.5~mag~arcsec$^{-2}$, similar to that obtained in the earlier study \citep{pun:2012}. However, this figure should be treated with care because the data were collected under vastly inhomogeneous conditions. Effects of factors such as time of observations, locations of study, astronomical factors, meteorological factors, and location-specific lighting usage patterns, could all contribute to the actual NSB observed. 
 
One of the main goals of our study was to quantify the skyglow caused by artificial lighting sources on the ground. As discussed earlier (cf Section~\ref{subsec:obssite}), the locations were broadly classified into urban and rural to characterize the effects of human activities. Combined statistics based on the location settings are presented in Table~\ref{tab:statistic-results}. As expected, all urban monitoring stations recorded brighter average night sky (henceforth a lower NSB value in mag~arcsec$^{-2}$ unit) than every rural station. Averaging all the data from the 10 urban stations and the 6 rural stations, it was found that the night sky in an urban setting (15.9~mag~arcsec$^{-2}$) was on average 2.5~mag, or 10 times\footnote{A difference of 1 mag arcsec$^{-2}$ refers to a light flux ratio of 2.512. Therefore a difference of $x$ mag arcsec$^{-2}$ refers to a flux ratio of $2.512^{x}$.}, brighter than in a rural one (18.4~mag~arcsec$^{-2}$). This again confirmed results from the previous study \citep{pun:2012} that human lighting contributed significantly to the light pollution conditions in Hong Kong. 

Even among the urban stations, there existed a large range of observed NSB, with the brightest site (TST at 14.8~mag~arcsec$^{-2}$) having a night sky 5 times brighter than the darkest urban site (TSW at 16.6~mag~arcsec$^{-2}$). This should not come as a surprise because of the large variations in the amount of external lighting used in the surroundings. In our case, the brightest urban location was the one situated in the middle of the commercial and business center in the city while the darkest urban location was located near the intersection between an ecological park and a town center (cf Section~\ref{subsec:obssite}).

\begin{figure}[ht]
\begin{center}
\includegraphics[width=8cm]{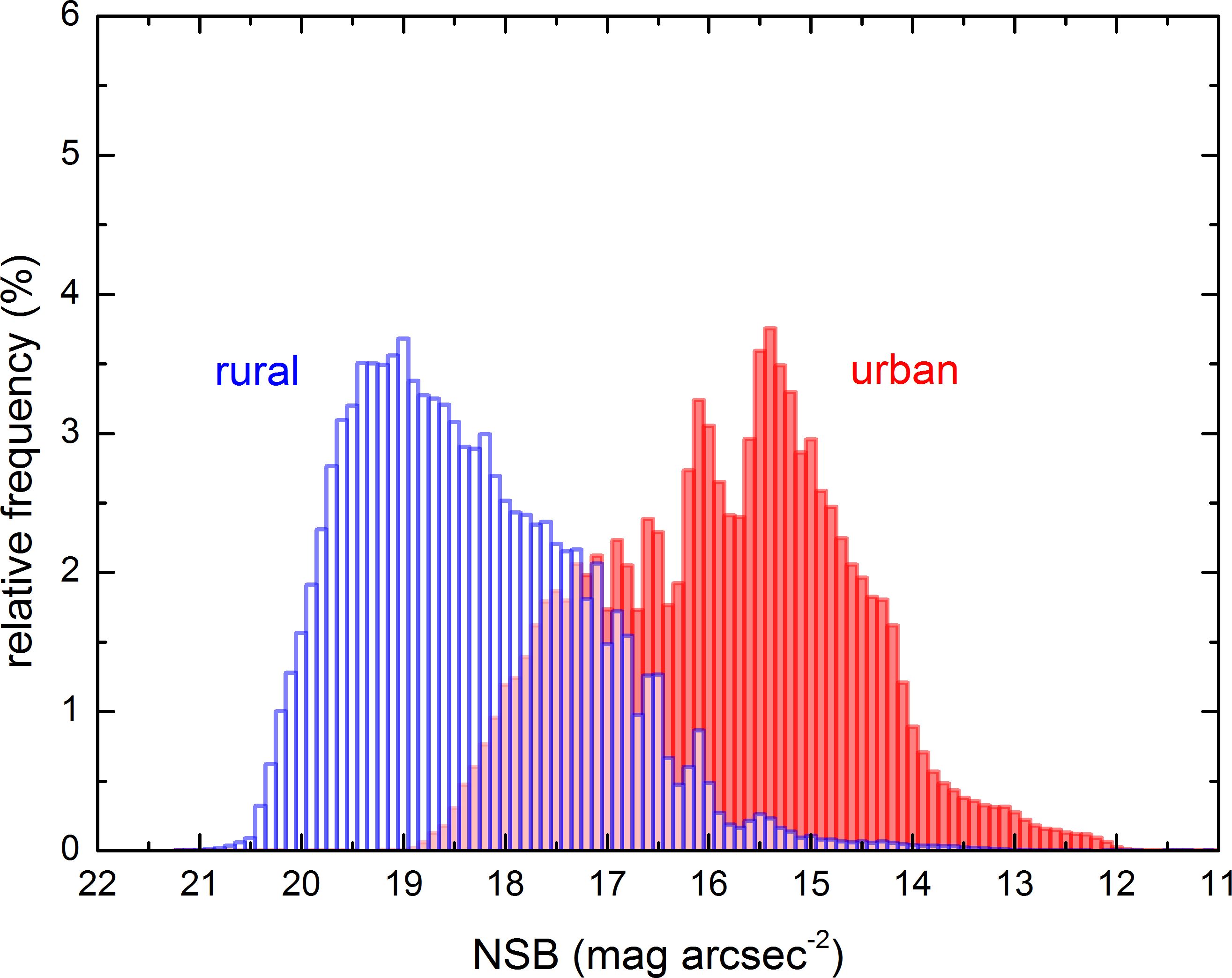}
\caption{Histogram showing the relative percentage distribution of all night sky brightness (NSB) recorded at the urban and rural stations. Different normalization factors were applied to the urban and rural data such that the number of urban and rural data bars will add up to 100\% separately.\label{fig:overall_histogram}}
\end{center}
\end{figure}

The combined histograms of NSB recorded at urban and rural stations were shown in Figure~\ref{fig:overall_histogram}. The relative frequencies of the urban and rural NSB data were computed separately to make up for the fewer stations and hence fewer NSB measurements received from the rural locations. The range of NSB values observed for each type of setting were large for either urban or rural-type locations, reflecting the huge inhomogeneities of the conditions under which these data were taken. NSB measurements taken from both types of settings could span a range of $\sim$~7~mag difference in brightness, or over 600 times in observed luminosity. On the other hand, it was clear from the figure that skyglow from urban lighting sources brightened up the night sky significantly, as witnessed from the split distributions of the urban and rural data taken under identical lunar and similar atmospheric conditions.


%


\subsection{Effects of moonlight on NSB distribution}\label{subsec:moon_nsb_distribution}

After effects of the Sun were removed in our database, the Moon remained the single biggest natural factor that affected the NSB observed. The amount of Moon's direct and scattered radiation at the zenith depended on both its altitude and phase which were calculated by the formulations in \citet{duffett:1981}. Roughly half of data were collected when the Moon was above the horizon. The SQM-LE recorded the combined radiation from the Moon in addition to scattered human light sources in these cases. Our dataset covers the entire physically possible ranges of Moon altitude and phase.


To study the effects of moonlight on the observed NSB, we adopted the moonlight brightness model as outlined in \citet{krisciunas:1991}. The model calculated the radiation from the scattered moonlight as a function of Moon phase and the angular separation between the Moon and the sky being observed, which was the zenith in our case. The model calculations were originally formulated for the astronomical broadband $V$ filter bandwidth (central wavelength 545 nm, FWHM 84 nm \citep{bessell:2005}), but had been directly adopted for the SQM-LE for simplicity. 



The observed change in the zenith NSB due to moonlight, $\Delta V_{\rm moon}$, was derived by comparing the modeled Moon's surface brightness and the zenith NSB background level, $B_0$. We adopted the darkest recorded NSB in the database, 21.2 mag arcsec$^{-2}$ measured at the AP station, as the baseline $B_0$. This would provide an upper limit to the effects from the Moon to the observed NSB in the entire study. Another parameter in the model was the atmospheric extinction coefficient $k$ for the $V$ wavelength band. The value of $k = 0.58$~mag/airmass, derived from a variable star observing project \citep{wayne:2011} performed at the astronomical telescope housed in the same building of the iObs station in October 2010, was adopted. 

\begin{figure*}[ht]
\begin{center}
\includegraphics[width=14cm]{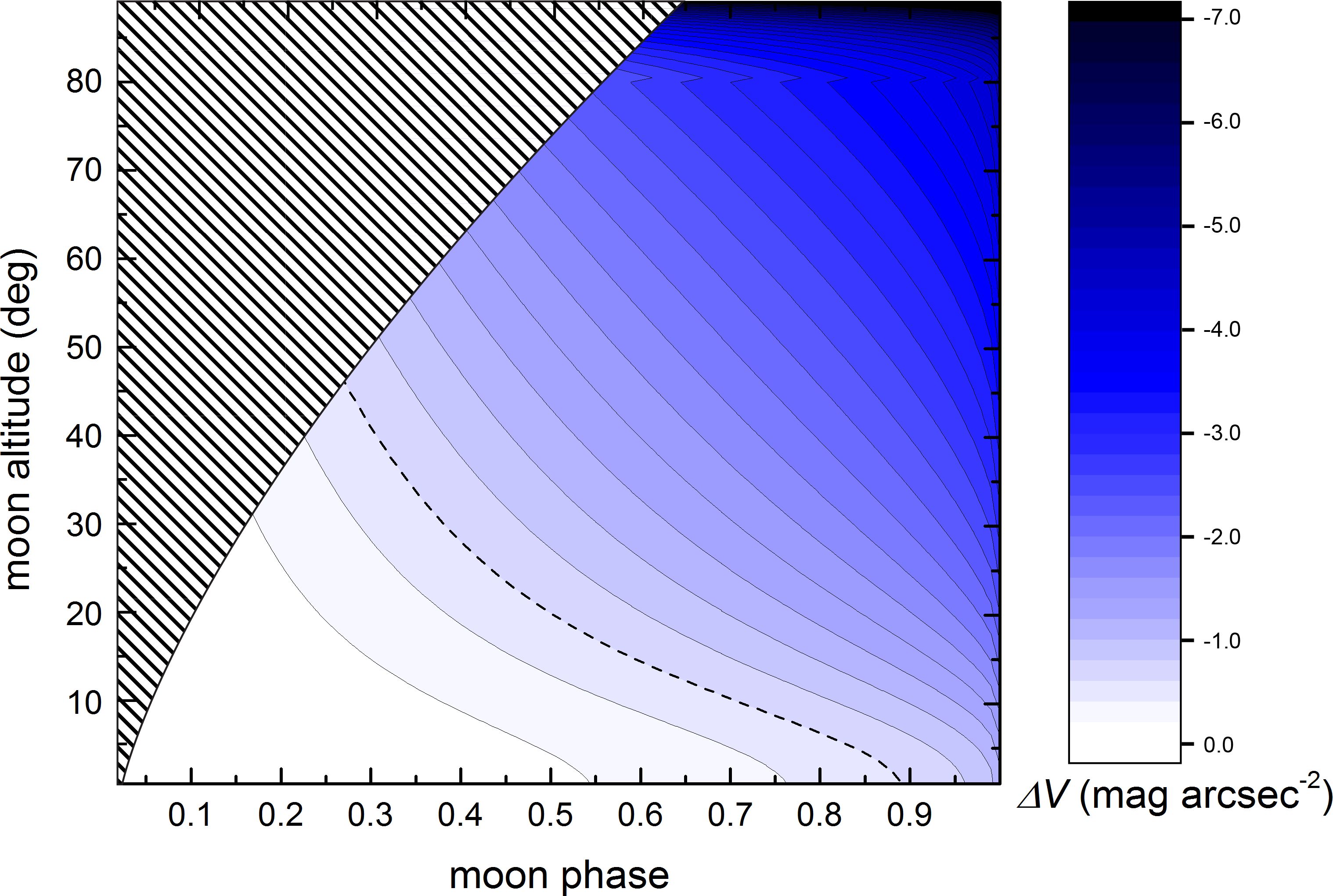}
\caption{Effect of moonlight on the zenith night sky brightness as a function of Moon's altitude and phase calculated from the moonlight prediction model of \citet{krisciunas:1991}. The $V$-band atmospheric extinction coefficient $k$ and the dark night-time zenith sky background level $B_0$ were assumed to be 0.58 mag/airmass and 21.2 mag arcsec$^{-2}$, respectively. The \textit{dashed} curve represents combinations of Moon altitude and Moon phase which gave $\Delta V_{\rm moon}$ = -0.6 mag arcsec$^{-2}$. The discontinuous contour lines at Moon altitude of 80$^{\circ}$ originated from the discontinuity of the scattering functions in Equations (18) and (19) of \citet{krisciunas:1991}. The top left \textit{diagonally shaded} region represented a region that was excluded for lunar motions during night-time.
\label{fig:ks91}}
\end{center}
\end{figure*}

The calculated $\Delta V_{\rm moon}$ contours for various Moon altitude and Moon phase combinations were shown in Figure~\ref{fig:ks91}. As expected, the values of $\Delta V_{\rm moon}$ were all negative, meaning that the net effect of the Moon would lead to a decrease in the observed NSB in mag~arcsec$^{-2}$ unit. The discontinuous contour lines at Moon altitude of 80$^{\circ}$ originated from the discontinuity of the scattering functions in Equations (18) and (19) of \citet{krisciunas:1991}.

Using these results, we adopted the value of $\Delta V_{\rm moon} = -0.6$ mag~arcsec$^{-2}$ as the line separating data that were affected by the Moon. NSB observations taken under Moon altitude-phase combinations such that $\Delta V_{\rm moon} > -0.6$ mag~arcsec$^{-2}$ would be classified as moonlight-free while the others would be classified as moonlight-affected. The value of 0.6~mag~arcsec$^{-2}$ was half of the standard deviation of NSB at the darkest station AP (Table~\ref{tab:statistic-results}). It was assumed that for any moonlight contribution at less than half of the observed spread in the NSB, the moonlight effect could not be easily distinguished from contributions from other factors such as variations of the atmospheric conditions.
   
It should be noted that in the current calculation of $\Delta V_{\rm moon}$, the angular response of the SQM-LE (Section~\ref{subsec:sqmle}) had not been included. If that were included, the effect from the Moon would be even bigger when the moonlight observed within the entire FOV of SQM-LE were integrated. In addition, the difference in spectral response between the SQM-LE sensor and photometric broadband $V$ filter, variations of the atmospheric conditions among individual observations (reflected by $k$), varying levels of moonless NSB in different stations (reflected by $B_0$), and the model prediction accuracy (claimed by \citet{krisciunas:1991} to be of order 10\% to 20\%) limited the data selection accuracy. On the other hand, we were only looking for a fast algorithm to investigate the overall effects of moonlight on NSB and we believe the proposed method should be sufficient. 

All the NSB data collected were classified as either moonlight-affected or moonlight-free using the Moon altitude-phase criteria described above. In the end, 59\% of the NSB measurements were classified as moonlight-free. The statistics of the moonlight-free data subset for all monitoring stations were also presented in Table~\ref{tab:statistic-results}. As expected, the removal of the moonlight-affected data in our database led to an increase in the average NSB (that is, reduction in average sky brightness) for all stations. On the other hand, effects of moonlight to the observed NSB were smaller in urban versus rural stations, as witnessed from the smaller rise in the averaged NSB when the moonlight-affected data were removed ($0.1 - 0.3$ mag~arcsec$^{-2}$ for the urban stations versus $0.3 - 0.6$ mag~arcsec$^{-2}$ for the rural ones). Combining data from all stations, the removal of moonlight-affected data led to an increase in the average observed NSB by 0.1 and 0.6 mag~arcsec$^{-2}$ respectively, for urban and rural sites.

After accounting for the moonlight, the effect of human lighting on the NSB could be described more accurately by comparing the average moonlight-free NSB values from different monitoring stations. The brightest location (TST at 14.9 mag~arcsec$^{-2}$) was 4.4 mag~arcsec$^{-2}$, or 57 times, brighter than the darkest one (AP at 19.3 mag~arcsec$^{-2}$). While the instantaneous weather in Hong Kong could vary by a great deal in different locations, the overall atmospheric conditions experienced at all locations should be roughly the same over the long duration of the project. Therefore it could be concluded that the large difference in the NSB between the brightest and the darkest sites was entirely due to the different lighting environments at these two locations. When the moonlight-free NSB data from all the urban (1.77 million measurements) and rural (661k measurements) stations were combined, it was found that the urban sites were on average 3.0 mag~arcsec$^{-2}$, or 15 times, brighter than the rural ones. 

The overall average moonlight-free NSB in Hong Kong was 16.8 mag~arcsec$^{-2}$. This figure could be compared to the natural zenith $V$-band NSB level of 21.6 mag arcsec$^{-2}$ suggested by the International Astronomical Union (IAU) for a good astronomical site with no pollution from artificial lighting sources~\citep{smith:1979}. It should be noted that the spectral response of the SQM-LE (240 nm FWHM) was broader than that of the broadband $V$ filter (84 nm FWHM), and the peak wavelengths of these two were also slightly different (540 nm for SQM-LE versus 545 nm for the $V$-band). The average Hong Kong night sky was 4.8 mag~arcsec$^{-2}$, or 82 times, brighter than that standard. In the worst (TST) and the least (AP) light-polluted locations, even after the effects of the Moon were mostly removed, the night sky was on average 464 and 8 times brighter than the standard respectively. On the other hand, the combined urban and rural averages, at 16.0 and 19.0 mag~arcsec$^{-2}$ respectively, were 168 and 11 times brighter than the standard. 

\begin{figure*}[ht]
\begin{center}
\includegraphics[width=16cm]{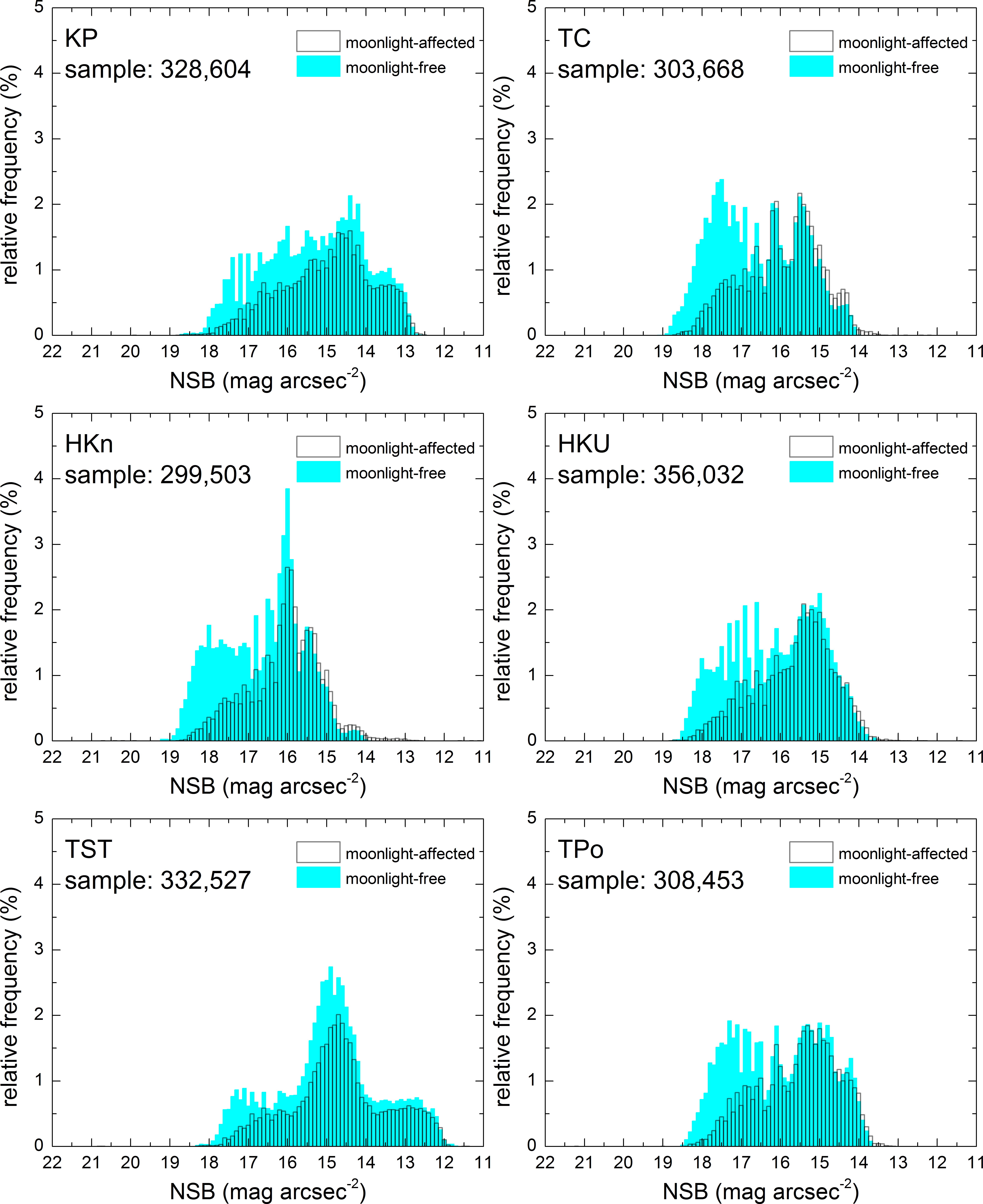}
\caption{Histograms showing the relative percentage distribution of the night sky brightness (NSB) recorded at the urban stations KP, TC, HKn, HKU, TST, and TPo, with (\textit{open bars}) and without (\textit{solid cyan bars}) the influence of moonlight. Fractions of the number of data against the total sample were calculated so that the number of moonlight-affected and moonlight-free data bars will add up to 100\%. Refer to Section~\ref{subsec:moon_nsb_distribution} for conditions of the moonlight-affected and moonlight-free data.   \label{fig:sitely_moon_histograms_urban1}}
\end{center}
\end{figure*}

\begin{figure*}[ht]
\begin{center}
\includegraphics[width=16cm]{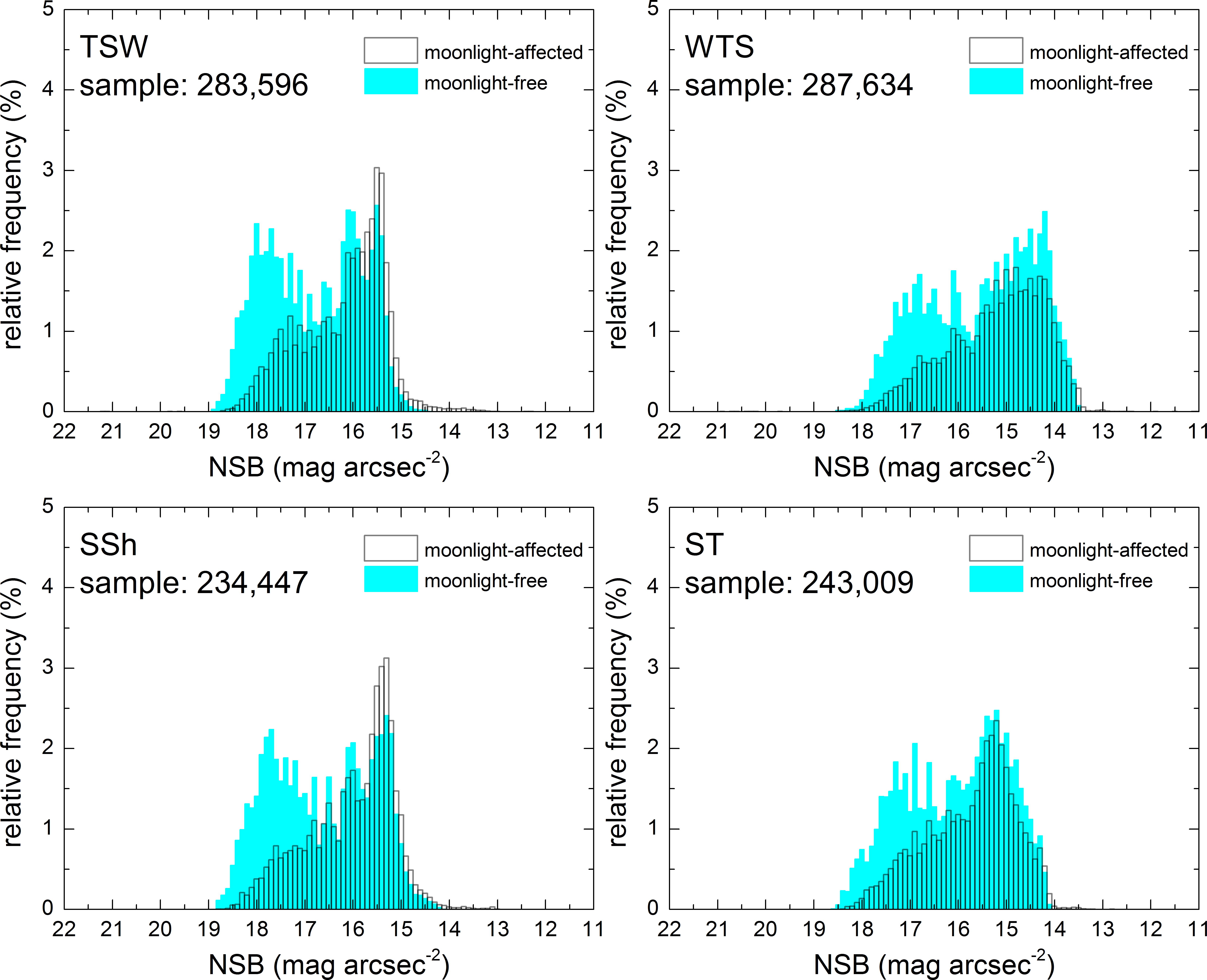}
\caption{Same as Figure~\ref{fig:sitely_moon_histograms_urban1} for the urban stations TSW, WTS, SSh, and ST. \label{fig:sitely_moon_histograms_urban2}}
\end{center}
\end{figure*}

\begin{figure*}[ht]
\begin{center}
\includegraphics[width=16cm]{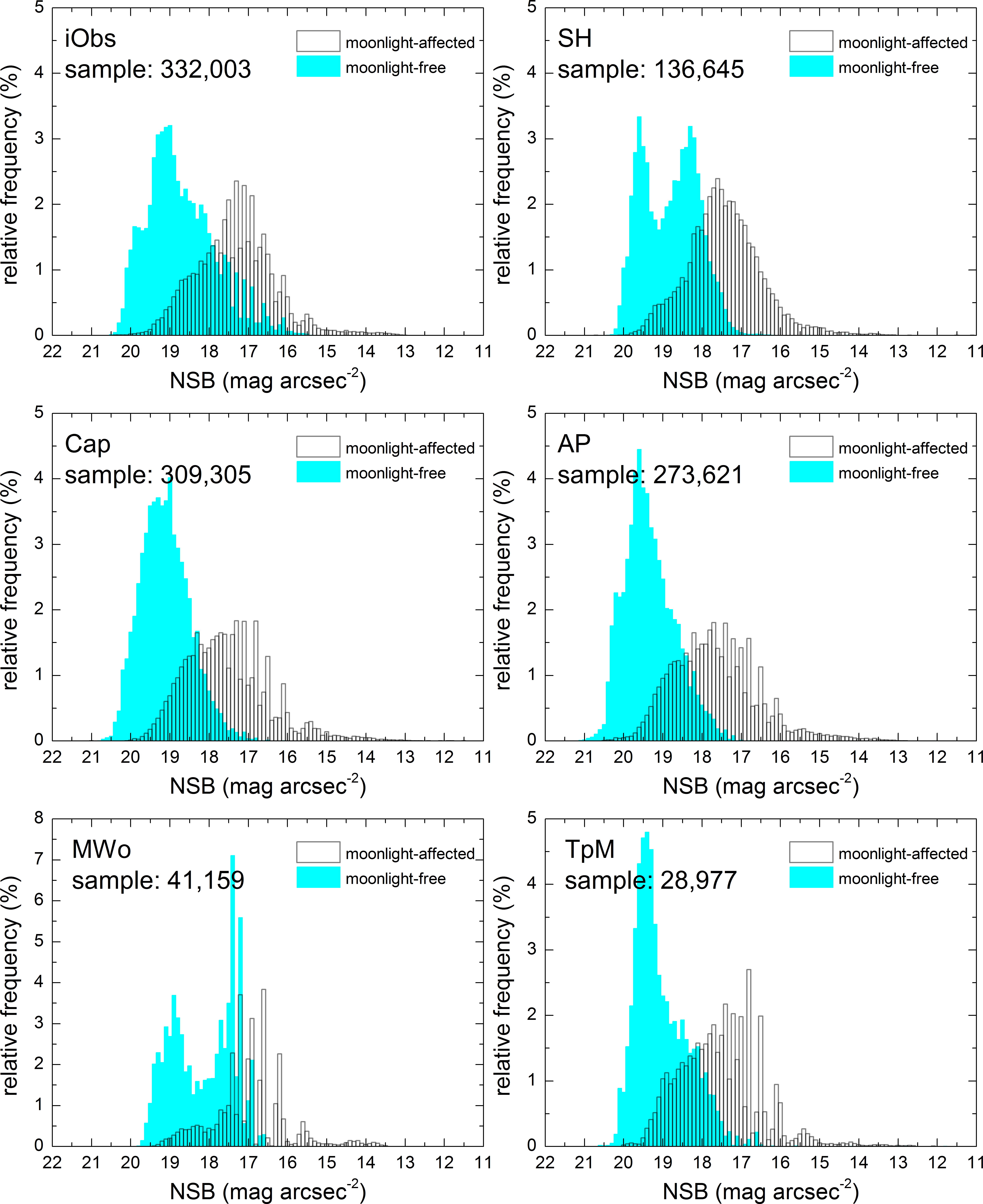}
\caption{Same as Figure~\ref{fig:sitely_moon_histograms_urban1} for the rural stations iObs, SH, Cap, AP, MWo, and TpM. Note that the number of measurements were lower for the MWo and TpM stations. Refer to Sections~\ref{subsec:obssite} and~\ref{subsec:data_flow} for reasons.
\label{fig:sitely_moon_histograms_rural}}
\end{center}
\end{figure*}

\begin{figure*}[ht]
\begin{center}
\includegraphics[width=16cm]{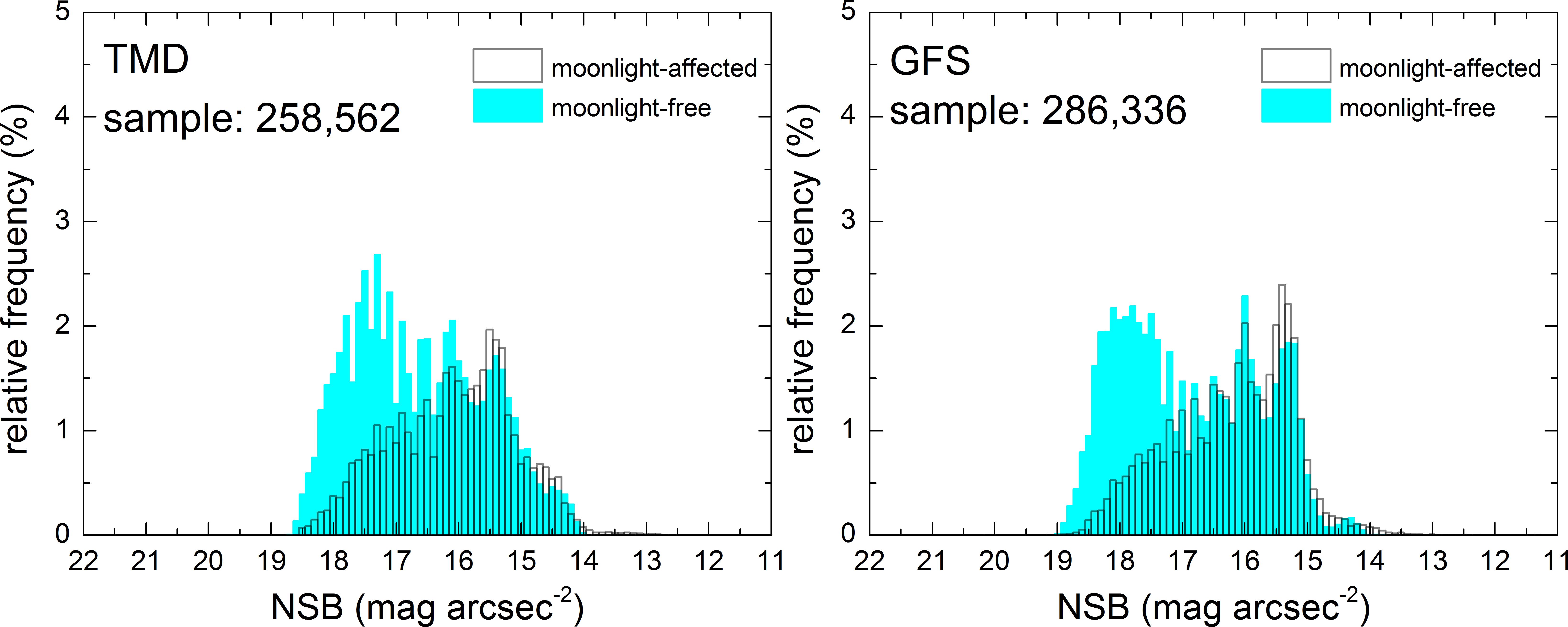}
\caption{Same as Figure~\ref{fig:sitely_moon_histograms_urban1} for the two stations TMD and GFS that were classified as neither urban nor rural. \label{fig:sitely_moon_histograms_not_classified}}
\end{center}
\end{figure*}

The histograms of NSB distributions for all monitoring stations are presented in Figures~\ref{fig:sitely_moon_histograms_urban1} to~\ref{fig:sitely_moon_histograms_not_classified} for different land settings. The relative frequencies compared to the total number of data points collected at that station are plotted. To illustrate the effects of moonlight on the observed NSB, the distributions from both the moonlight-affected and moonlight-free data were overlaid for comparison. As in the combined histogram in Figure~\ref{fig:overall_histogram}, the observed NSB at each site spanned a large range in brightness, with a typical range of $\sim 5 - 6$ mag~arcsec$^{-2}$ for an urban or an unclassified location, and $\sim 6 - 7$ mag~arcsec$^{-2}$ for a rural site. The distribution of NSB at any particular site could not be described by any simple function. Moreover, there was not a single characteristic shape to describe the NSB distributions in all monitoring stations. This inhomogeneity was most probably due to the different lighting environments near different stations. 

For most of the urban (all except KP and TST) and for the two unclassified stations, the distributions of the NSB near the bright end (small NSB values) of the histograms were similar irrespective of the presence of the Moon, while the moonlight-affected and moonlight-free distributions differed at the dim end (large NSB values), with an additional peak for the moonlight-free data. These observations suggested that when the night sky was bright at an urban site, emission from the Moon did not contribute much to the observed NSB. Other factors, such as the lighting environment, the time in the evening, and the cloud amount \citep{christopher:2011} were likely to be more important in dictating the observed NSB. On the other hand, when a dark sky was observed at an urban setting, it was most likely to be a moonlight-free evening. In KP and TST, the two brightest locations in our network, the shapes of the moonlight-affected and moonlight-free histograms were similar at either the bright or dim end of the plot. This implied that radiation of the Moon contributed only a small fraction of the observed NSB at all times, thus resulting in similar distributions observed under all moonlight conditions. 

For the rural stations, the separation of the observed NSB based on the amount of moonlight resulted in two distinct populations (Figure~\ref{fig:sitely_moon_histograms_rural}), with different peaks observed in the moonlight-affected and moonlight-free data subsets. This indicated that emission from the Moon was a dominating light source at these rural settings whenever it satisfied the moonlight-affected conditions. Another conclusion is that the amount of the artificial lighting used in these environments was much lower than in the urban environments. It is believed that other factors, such as the varying cloud amount, contributed to the spread of the moonlight-free NSB observed. The removal of moonlight-affected data also led to a reduction in the range of moonlight-free NSB data to only $\sim$~4 mag~arcsec$^{-2}$, as well as a smaller standard deviation as listed in Table~\ref{tab:statistic-results} (from $1.0 - 1.2$ mag~arcsec$^{-2}$ for all data to $0.6 - 0.9$ mag~arcsec$^{-2}$ for moonlight-free data). On the other hand, the spread of the urban NSB data stayed fairly constant even as we studied the moonlight-free data subset. 

As seen in Figure~\ref{fig:sitely_moon_histograms_rural}, several types of distributions exist for the moonlight-free NSB distribution in rural stations. For the Cap and AP stations, the moonlight-free data were roughly in Gaussian distributions, with slight extension towards the brighter end. On the other hand, in the SH and MWo stations, both with similar rural village environments, two peaks separated by $\sim 1.5 - 2$~mag~arcsec$^{-2}$ are observed. It should be noticed though that data taken at the MWo station suffered from the digitization problem in the beginning of the survey, resulting in some alternating distribution pattern for data brighter than 17.0~mag~arcsec$^{-2}$ (cf Section~\ref{subsec:sqmle}). Finally, some hybrid distributions seem to be present for the iObs and TpM data. It was interesting to notice that similar double-peak distributions can also be found in most of the urban moonlight-free histograms in Figures~\ref{fig:sitely_moon_histograms_urban1} and~\ref{fig:sitely_moon_histograms_urban2}. The observed double-peak pattern probably resulted from the combined effects of different factors, including the variations of the NSB as a function of time, season, and cloud amount. The last factor had been investigated earlier in studies such as \citet{christopher:2011} in which the cloud coverage could brighten the observed sky brightness due to enhanced back-scattering of artificial light.

\begin{figure}[ht]
\begin{center}
\includegraphics[width=8cm]{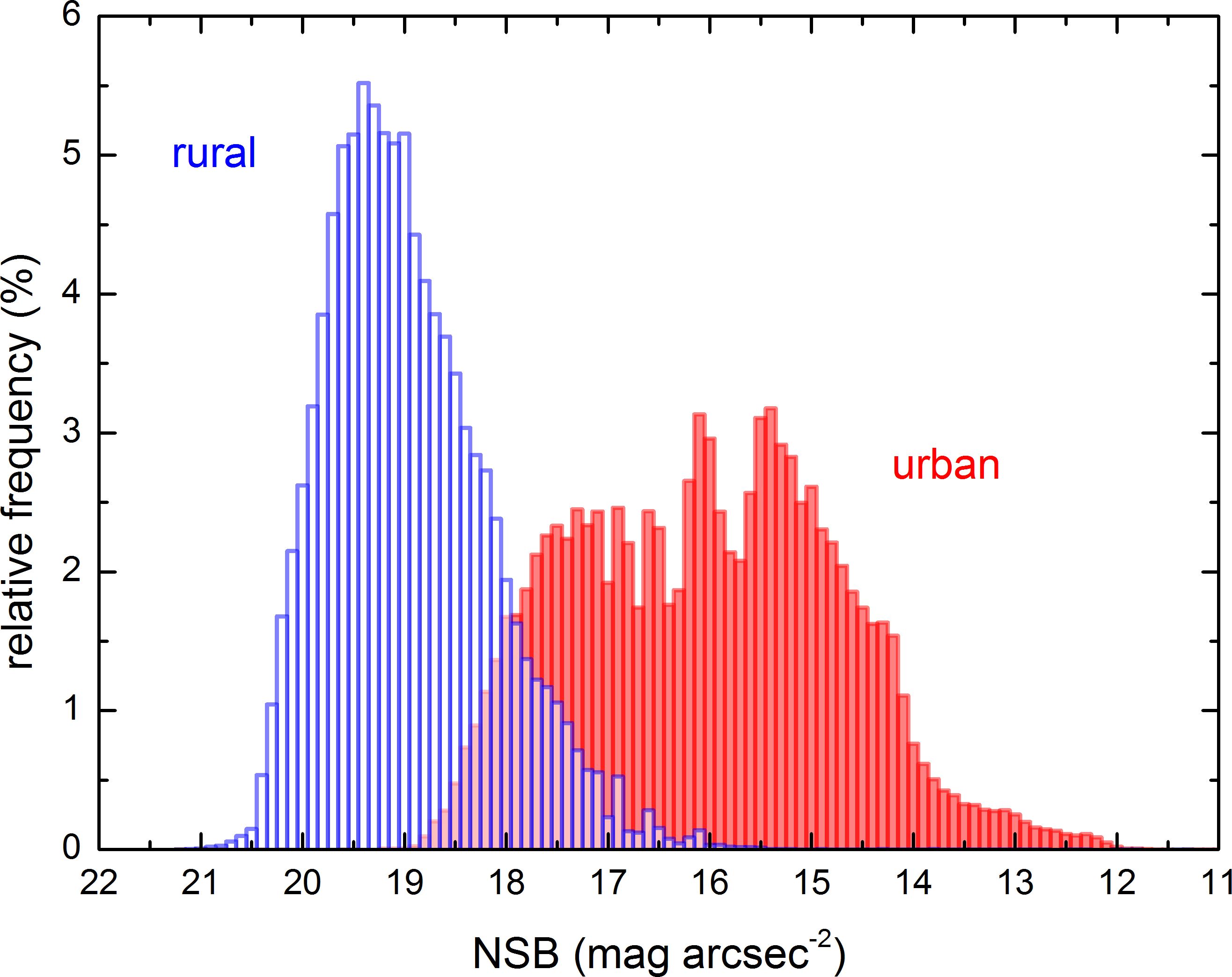}
\caption{Histogram showing the relative percentage distribution of moonlight-free night sky brightness (NSB) recorded at urban and rural stations. Different normalization factors were applied to urban and rural data, such that the number of urban and rural data bars will add up to 100\% separately. Refer to Section~\ref{subsec:moon_nsb_distribution} for the selection of moonlight-free data. \label{fig:overall_moonless_histogram}}
\end{center}
\end{figure}

Finally, a comparison of the urban versus rural distribution for the moonlight-free NSB is presented in Figure~\ref{fig:overall_moonless_histogram}. This plot is similar to the histogram in Figure~\ref{fig:overall_histogram}, with the relative frequencies of the urban and rural NSB data were computed separately. 
After removing the moonlight-affected data, both the urban and rural distributions exhibit larger occurrence near the dark side as expected. 
Moreover, both histograms in Figure~\ref{fig:overall_moonless_histogram} skewed towards the brighter end (smaller values of NSB), which could possibly be explained by the skyglow caused by clouds in an overcast night sky. 
The average difference between the moonlight-free urban and rural NSB was 3.0~mag arcsec$^{-2}$, or a flux ratio of $\sim$~15 times in light intensity. Although the effects of variations in NSB due to other factors such as the cloud amount have not been removed, this difference represents a characteristic number when we contrast the quality of the night sky between urban and rural settings in Hong Kong. 

%% file: table-statistic.tex
\begin{table*}[ht]														
\caption{Statistics of the NSB measurements in Hong Kong. Refer to Section~\ref{subsec:moon_nsb_distribution} for the definition of the moonlight-free condition.}
\centering 														
\begin{tabular}{c r c c r c c}														
\hline\hline 														
														
	&	\multicolumn{3}{c}{\textbf{All data}} 					&	\multicolumn{3}{c}{\textbf{Moonlight-free data}} 					 \\ 	
Station	&	Sample size	&    	\multicolumn{2}{c}{NSB in mag arcsec$^{-2}$} 			&	Sample size	&    	\multicolumn{2}{c}{NSB in mag arcsec$^{-2}$} 			 \\ 	\cline{3-4} \cline{6-7}
 (no. of stations)	&		&	Average	&	Standard deviation	&		&	Average	&	Standard deviation	 \\ 	
\hline														
all (18)           & 4,644,081 & 16.5 & 1.6 & 2,758,461 & 16.8 & 1.7 \\
Urban (10)         & 2,977,473 & 15.9 & 1.3 & 1,771,888 & 16.0 & 1.3 \\
Rural (6)          & 1,121,710 & 18.4 & 1.2 & 660,931   & 19.0 & 0.8 \\
Not classified (2) & 544,898   & 16.5 & 1.1 & 325,642   & 16.7 & 1.1 \\
														
\hline														
\multicolumn{7}{c}{\textbf{Urban}}	 \\													
KP                 & 328,604   & 15.2 & 1.3 & 199,389   & 15.3 & 1.3 \\
TC                 & 303,668   & 16.3 & 1.1 & 178,638   & 16.6 & 1.1 \\
HKn                & 299,503   & 16.4 & 1.0 & 181,860   & 16.7 & 1.1 \\
HKU                & 356,032   & 15.9 & 1.1 & 210,993   & 16.1 & 1.2 \\
TST                & 332,527   & 14.8 & 1.3 & 198,033   & 14.9 & 1.3 \\
TPo                & 308,453   & 15.8 & 1.1 & 181,818   & 16.0 & 1.2 \\
TSW            & 283,596   & 16.6 & 1.0 & 165,344   & 16.8 & 1.0 \\
WTS                & 287,634   & 15.4 & 1.1 & 174,689   & 15.6 & 1.2 \\
SSh                & 234,447   & 16.4 & 1.0 & 135,047   & 16.6 & 1.1 \\
ST                 & 243,009   & 15.9 & 1.0 & 146,077   & 16.1 & 1.1 \\
														
\hline														
\multicolumn{7}{c}{\textbf{Rural}}	 \\													
iObs               & 332,003   & 18.1 & 1.1 & 195,456   & 18.7 & 0.9 \\
SH                 & 136,645   & 18.2 & 1.1 & 74,977    & 18.8 & 0.7 \\
Cap                & 309,305   & 18.5 & 1.1 & 184,677   & 19.1 & 0.6 \\
AP                 & 273,621   & 18.7 & 1.2 & 162,011   & 19.3 & 0.7 \\
MWo                & 41,159    & 17.8 & 1.0 & 27,306    & 18.1 & 0.8 \\
TpM                & 28,977    & 18.4 & 1.1 & 16,504    & 19.0 & 0.7 \\
														
\hline														
\multicolumn{7}{c}{\textbf{Not classified}}	 \\													
TMD                & 258,562   & 16.4 & 1.1 & 156,508   & 16.6 & 1.1 \\
GFS                & 286,336   & 16.6 & 1.1 & 169,134   & 16.9 & 1.1 \\
\hline														
\end{tabular}														
\label{tab:statistic-results} 														
\end{table*}

%% file: discussions.tex
Outdoor lighting is an integral and indispensible part of modern societies. The accompanying issue of light pollution has just begun to arouse the attention of a wide spectrum of the general public, as well as ecologists, medical professionals, astronomers, and energy conservationists. 
To combat this growing environmental problem, a Global Project titled \textit{Dark Skies Awareness}\footnote{See \url{http://www.astronomy2009.org/globalprojects/cornerstones/darkskiesawareness/}} was launched during the International Year of Astronomy 2009. Studies of the night sky brightness (NSB), long the exclusive domain of professional astronomers, provide a measure of the extent of light pollution. While earlier measurements of the NSB were mostly conducted at dark astronomical observing sites, a NSB study in an urban environment allows the impacts of outdoor lighting fixtures in various configurations to be investigated. Using a portable and easy-to-use solid-state measuring device called the Sky Quality Meter (SQM), we have previously conducted a study in Hong Kong, collecting over 2,000 NSB measurements over a 18 month duration in 2008-09 \citep{pun:2012}. Results of this path-finding study clearly revealed that human outdoor lighting usage contributed strongly to this environmental degradation. 

Building upon the earlier survey, we established a more comprehensive monitoring of the night sky by setting up the \textit{Hong Kong Night Sky Brightness Monitoring Network} (NSN) in May 2010, which continually measured the NSB at multiple urban and rural sites. Real-time maps were generated to depict the night sky conditions in Hong Kong using the ethernet version of the SQM, known as Sky Quality Meter - Lens Ethernet (SQM-LE). A total of 18 stations were established to monitor the variations of the NSB in Hong Kong due to large variations in the landscape and urban developments in the city. By choosing these monitoring locations strategically over different regions with diverse land uses around Hong Kong, the NSN aimed to monitor changes in NSB automatically as a function of time, location, and others environmental factors. The installation of each station was designed to be low cost, weatherproof, and flexible. The operation of each station was designed to be robust and maintenance free. Real-time data were obtained from the remote sites via the mobile Internet network. The end-product was a huge dataset with high temporal resolution collected simultaneously at multiple locations. Over 4.6 million individual NSB measurements were recorded under a huge range of environmental conditions over 35 months since the start of the project. This was over 2,000 times more data collected than the previous study and represented one of the largest NSB data sets heretofore collected for a single city. This formed the basis for a comprehensive investigation of the underlying relations between the NSB and various artificial and natural factors. 

The overall average brightness level of the Hong Kong night sky, after removing the data taken during moonlight-affected conditions, was 16.8 mag arcsec$^{-2}$, which was almost 100 times higher than that of the pristine night sky background as defined by the International Astronomical Union \citep{smith:1979}. From a rough classification based on satellite imagery, land use, and population information, moonlight-free observations made in an urban night sky is on average 3.0 mag arcsec$^{-2}$, or 15 times, brighter in light intensity than that in rural locations. While this result suffered from the inherent diversity of broad concepts such as urban and rural, it firmly established the level of influence of human lighting on the environment. The general and broad nature of the classification was also reflected in the large variations of average NSB observed among stations of either type of setting, suggesting diverse ambient artificial lighting present. Finally, the limited number of observing stations was insufficient to cover all the variations in NSB in a geographically complex metropolis such as Hong Kong. A more systematic design of the monitoring network, such as arranging for monitoring stations based on population distribution, might be required to determine a more precise picture of the exact effects of urban lighting. On the other hand, it is difficult to draw any straight-forward conclusion from this study on how the observed NSB changes with distance from the center of a large city. It is because, unlike some other cities where there is a single city center radially spread out in all directions on a flat landscape, urban settlements in Hong Kong are distributed over large areas with complex terrain and hilly landscape (cf Figure~\ref{fig:station-map}). 

It is worthwhile to compare the findings on the NSB properties of Hong Kong from this study with those collected elsewhere. During 2012~-- 2013, another survey based on data collected by two SQM-LE units in two locations (one urban and one rural) was conducted in Vienna, Austria~\citep{johannes:2013}. Instead of quoting an overall average, the authors reported typical values of NSB (regardless of Moon conditions) in urban Vienna of 16.3 and 19.1 mag~arcsec$^{-2}$ during overcast and clear conditions respectively. While direct comparisons between the results from the two cities were not possible (it was not clear in \citet{johannes:2013} the exact quantitative criteria of the classification of overcast and clear night skies), it was noticed that only $\sim$0.007\% of urban data in Hong Kong (cf distribution of urban NSB in Figure~\ref{fig:overall_histogram}) was darker than the \textit{typical} NSB measured in Vienna in clear conditions. While lacking detailed meteorological data in our survey, this result appears to argue for a more severe light pollution problem in Hong Kong. This is consistent with our conclusion that man-made lighting was the main pollution source in urban areas as the population density in Hong Kong (6,620 people per km$^2$) is higher than that in Vienna (4,002 people per km$^2$).

The major challenge for establishing this long-term data archive on the NSB was to ensure the quality and consistency of data collected. The main concern was the consistency of all the optical components over the project duration, including the long-term stability of the SQM-LE and the transparency of the protective casing. These were addressed by regular servicing trips and calibrations performed using the laboratory test-stand to check the performance of each optical component. Under the constraints of limited budget and manpower, they were done in a frequency of roughly once every $6 - 9$ months. This was believed to be enough to account for long-term gradual changes such as the transparency of the casing and the responses of most SQM-LE units. On the other hand, the current setup had difficulties tracking sudden drops in performance (such as the frosted filter glasses of two SQM-LE units, cf Section~\ref{subsec:data_quality_calibration}) until after the problem was discovered during servicing. In summary, we believed the overall accuracy of the data in the archive would be of order $\sim 0.1 - 0.2$ mag arcsec$^{-2}$, of which 0.1 mag was the intrinsic accuracy of the SQM-LE claimed by the manufacturer. It should be noted that the effects of almost all these processes, such as aging of sensor, aging of casing, or dust, would have a net effect of reducing the brightness of the night sky observed. Therefore the night sky flux needs to be adjusted upwards, or decreases the NSB measured in mag arcsec$^{-2}$ unit, from what was in the archive. 

The present analysis has just scratched the surface of the huge database assembled. Further analyses of the data collected could provide valuable insights on how various natural and artificial factors affect the final NSB observed. Topics that our team is currently working on include detailed geographic (population, landscape, and land-usage) and temporal (seasonal as well as nightly) variations of the NSB, studies of relations between NSB and the Moon, NSB and the cloud (total amount, base height, and type), and NSB observed versus remote sensing data. Furthermore, as the NSB was one of the indicator of the extent of light pollution, extending the time baseline of this archive could yield the long-term trend of light pollution conditions in Hong Kong. It is expected that findings from these studies would contribute to the overall environment of Hong Kong by promoting light pollution reduction and energy saving. This monitoring network could possibly serve as the prototype for a future system with which the quality of night sky would be comprehensively monitored similar as other environmental indicators such as air quality and particulate concentrations.

%% file: acknowledge.tex
The \textit{Hong Kong Night Sky Brightness Monitoring Network} was supported by the Environment and Conservation Fund (Project ID: 2009-10) of the Hong Kong SAR government. 
We would like to thank three reviewers who read the manuscript in great detail, along with their encouraging remarks and valuable comments. We would also like to thank the co-organizers, namely, Hong Kong Observatory (HKO), Hong Kong Space Museum (HKSpM), Hong Kong Astronomical Society (HKAS), Ho Koon Nature Education cum Astronomical Centre (Ho Koon), and The Camping Association of Hong Kong, China, Ltd for continuous support on this project, particularly Siu-wai Chan, Ying-wah Chan, Ping-wah Li, Hing-yim Mok, Kit-chi Tsui, and Wai-kin Wong of HKO, Ki-hung Chan, Man-hung Ho, and Kai-ip Lau of HKSpM, Ian Chung and Bill Yeung of HKAS, and Kenneith Hui of Ho Koon, for their help. Officers and staff in the HKU Swire Institute of Marine Science, Environmental Protection Department, Agriculture, Fisheries and Conservation Department, Government Flying Service, Hong Kong Science Museum, Hong Kong Playground Association Silvermine Bay Outdoor Recreation Camp, and teachers in Our Lady's College, Elegantia College (Sponsored By Education Convergence), POH Chan Kai Memorial College are also acknowledged their help in the construction of the monitoring stations. Last but not least, we thank Anthony Tekatch from Unihedron for comprehensive and speedy technical assistance.

